\input harvmac
\newcount\figno
\figno=0
\def\fig#1#2#3{
\par\begingroup\parindent=0pt\leftskip=1cm\rightskip=1cm\parindent=0pt
\global\advance\figno by 1
\midinsert
\epsfxsize=#3
\centerline{\epsfbox{#2}}
\vskip 12pt
{\bf Fig. \the\figno:} #1\par
\endinsert\endgroup\par
}
\def\figlabel#1{\xdef#1{\the\figno}}
\def\encadremath#1{\vbox{\hrule\hbox{\vrule\kern8pt\vbox{\kern8pt
\hbox{$\displaystyle #1$}\kern8pt}
\kern8pt\vrule}\hrule}}
\def\W{Y}
\overfullrule=0pt

%macros
%
\def\tilde{\widetilde}
\def\bar{\overline}
\def\Z{{\bf Z}}

\def\S{{\bf S}}
\def\R{{\bf R}}

\font\zfont = cmss10 %scaled \magstep1

\def\bigone{\hbox{1\kern -.23em {\rm l}}}
\def\ZZ{\hbox{\zfont Z\kern-.4emZ}}

\Title{hep-th/9706109, IASSNS-HEP-97-66}
{\vbox{\centerline{BRANES AND THE DYNAMICS OF QCD}
\bigskip}}
\smallskip
\centerline{Edward Witten\foot{Research supported in part
by NSF  Grant  PHY-9513835.}}
\smallskip
\centerline{\it School of Natural Sciences, Institute for Advanced Study}
\centerline{\it Olden Lane, Princeton, NJ 08540, USA}\bigskip

\medskip

\noindent

A brane configuration is described that is relevant to understanding the 
dynamics
of $N=1$ supersymmetric Yang-Mills theory.  Confinement and spontaneous
breaking of a discrete chiral symmetry
 can be understood as consequences of the topology of the brane.
Because of the symmetry  breaking, there
can be domain walls separating different vacua; the QCD string can end on such
a domain wall.  The model in which these properties can be understood 
semiclassically
does not coincide with supersymmetric Yang-Mills theory but is evidently
in the same universality class.

\Date{June, 1997}
%text of paper

\newsec{Introduction}

Lately, many models of supersymmetric 
gauge theory in different dimensions have been studied
using various approaches based on string theory.
These models have varying applications and shed
light on many different aspects of strongly interacting gauge theories.  The 
purpose of the present paper is to apply some of these methods to the 
most classical riddles of four-dimensional gauge dynamics --  such as
confinement, chiral symmetry breaking,
and the relation, if any, between them -- in a situation as realistic as 
possible
while still more or less tractable.

To this aim, we want to explore
a supersymmetric model that is a relatively close
analog of ordinary non-supersymmetric QCD.  Such a model
is the minimal $N=1$ model with an $SU(n)$ vector multiplet only and no
chiral superfields.  This model is believed to exhibit confinement, a mass gap,
and spontaneous breaking of a (discrete) chiral symmetry.  These are some
of the main properties that we would like to understand in ordinary QCD.
Many other interesting supersymmetric models contain massless scalars,
as a result of which they are not such close relatives of QCD.

\nref\thooft{G. 't Hooft, ``A Planar Diagram Theory For Strong Interactions,''
Nucl. Phys. {\bf B72} (1974) 461.}
\nref\witten{E. Witten, ``Baryons In The $1/N$ Expansion,''Nucl. Phys. 
{\bf B160} (1979) 57.}
\nref\douglas{M. R. Douglas and S. H. Shenker,
``Dynamics Of $SU(N)$ Supersymmetric Gauge Theory,'' Nucl. Phys. {\bf B447} 
(1995)
271.}
\nref\bfss{T. Banks, W. Fischler, S. H. Shenker,
and L. Susskind, ``$M$-Theory As A Matrix Model: A Conjecture,''
Phys. Rev. {\bf D55} (1997) 5112.}
Another important property that the minimal $N=1$ model is believed to share
in common with ordinary QCD is a large $n$ limit in which confinement and
the mass gap persist, hadron masses approach fixed limits, and the residual
interactions among hadrons vanish \refs{\thooft,\witten}.  
Understanding of this limit, perhaps via a new kind of string theory, is 
probably
well out of reach with present methods, but may offer the best long term hope
of a much better understanding of QCD than we now possess.
Note that $N=2$ supersymmetric Yang-Mills theory is not believed to
have a large $n$ limit of the conventional sort \douglas; this fact
is important in the matrix model of $M$-theory \bfss.

\nref\hanany{A. Hanany and E. Witten,
``Type IIB Superstrings, BPS Monopoles, and Three-Dimensional Gauge Dynamics,''
Nucl. Phys. {\bf B492} (1997) 152.}
\nref\berkeley{ J. de Boer, K. Hori, Y. Oz, and Z. Yin,
``Branes And Mirror Symmetry In $N=2$ Supersymmetric Gauge Theories In
Three-Dimensions,'' hep-th/9702154.}
\nref\elitzur{
S. Elitzur, A. Giveon, and D. Kutasov, ``Branes And $N=1$ Duality In
String Theory,'' hep-th/9702014.}
\nref\elitzurtwo{ S. Elitzur, A. Giveon, D. Kutasov, E. Rabinovici, and
A. Schwimmer, ``Brane Dynamics And $N=1$ Supersymmetric Gauge Theory,''
hep-th/9704104.}
\nref\morewitten{E. Witten, ``Solutions Of Four-Dimensional Field
Theories From $M$-Theory,'' hep-th/9703166.}
\nref\evans{N. Evans, C. V. Johnson, A. D. Shapere, ``Orientifolds, Branes,
And Duality of 4D Gauge Theories,'' hep-th/9703210.}
\nref\brodie{J. Brodie and A. Hanany, ``Type IIA Superstrings, Chiral
Symmetry, and $N=1$ 4D Gauge Theory Dualities,'' hep-th/9704043.}
\nref\brand{A. Brandhuber, J. Sonnenschein, S. Theiren, and S. Yankielowicz,
``Brane Configurations And 4D Field Theory Dualities,'' hep-th/9704044.}
\nref\ahar{O. Aharony and A. Hanany, ``Branes, Superpotentials, And 
Superconformal
Fixed Points,'' hep-th/9704170.}
\nref\tatar{R. Tatar, ``Dualities In 4D Theories With Product Gauge Groups
From Brane Configurations,'' hep-th/9704198.}
\nref\brunner{I. Brunner and A. Karch, ``Branes And Six-Dimensional Fixed 
Points,''
hep-th/9705022.}
\nref\kol{B. Kol, ``5D Field Theories And M Theory,'' hep-th/9705031.}
\nref\morozov{A. Marshakov, M. Martellini, and A. Morozov,
``Insights and Puzzles From Branes; 4d SUSY Yang-Mills From 6d Models,''
hep-th/9706050.}
\nref\landsteiner{K. Landsteiner, E. Lopez, and D. A. Lowe,
``$N=2$ Supersymmetric Gauge Theories, Branes, And Orientifolds,''
 hep-th/9705199. }
\nref\otheris{A. Brandhuber, J. Sonnenschein, S. Theisen, and S. Yankielowicz,
``Brane Configurations And $4-D$ Field Theory Dualities,''
hep-th/9704044, ``$M$ Theory And Seiberg-Witten Curves: Orthogonal
And Symplectic Groups,'' hep-th/9705232.}
To study the pure $N=1$ gauge theory, we follow a path that has been followed
for a variety of three, four, and five-dimensional models \refs{\hanany - 
\otheris}: we construct a configuration of branes in a weakly coupled
string theory that realizes the field theory of interest
at low energies, and then we solve the model in the strong coupling limit
by using some of the string theory dualities.
\foot{A quite different way to study gauge theory dynamics via string theory
involves encoding four-dimensional gauge dynamics in Type IIA and Type IIB
geometry near singularities, a process initiated in \ref\kv{S. Kachru and
C. Vafa, ``Exact Results For $N=2$ Compactifications Of Heterotic Strings,''
Nucl. Phys. {\bf B450} (1995) 69, 
hep-th/950510.} and developed subsequently
in many directions.  In \ref\warner{A. Klemm, W. Lerche, P. Mayr, C. Vafa,
and N. Warner, ``Self-Dual Strings And $N=2$ Supersymmetric Field Theory,''
hep-th/9604034}, such methods were used to propose a fivebrane interpretation
of some four-dimensional field theories, and in 
\ref\mayr{S. Katz, P. Mayr, and C. Vafa,
``Mirror Symmetry And Exact Solution Of 4D $N=2$ Gauge Theories, I''
 hep-th/9706110.} this was extended to give exact
exact solutions of a very large family of four-dimensional $N=2$ models.} 
  A method of doing this in the
case at hand will be described in section 2.  We analyze the symmetries of the
model and show that the ${\bf Z}_n$ chiral symmetry is spontaneously broken. 

Then in sections 3 and 4, we analyze further properties of the model.
In section 3, we show that the model is confining: it has strings or flux 
tubes,
which can terminate on an external quark, and can annihilate in groups of $n$.
The ability of $n$ identical strings to annihilate
reflects the fact that $n$ quarks -- that is $n$ copies of the fundamental
representation of $SU(n)$ -- can be combined to make an $SU(n)$-invariant 
``baryon.''

Chiral symmetry breaking means that the theory has $n$ distinct vacua, and 
therefore there are domain walls connecting them.  It has been argued
\ref\dvali{G. Dvali and M. Shifman,
``Domain Walls In Strongly Coupled Theories,'' Phys. Lett. {\bf B396} (1997) 
64,
hep-th/9612128; 
A. Kovner, M. Shifman, and A. Smilga, ``Domain Walls in Supersymmetric 
Yang-Mills Theories,'' 
hep-th/9706089.}  that the domain walls of this model are
BPS-saturated. This is also supported by the existence 
(see p. 635 of \ref\vafa{S. Cecotti and C. Vafa,
``On Classification of N=2 Supersymmetric Theories,'' Commun. Math. Phys.
 {\bf 158} (1993) 569.}) of BPS-saturated
domain walls in an effective field theory \ref\veneziano{ 
G. Veneziano and S. Yankielowicz, ``An Effective Lagrangian For The Pure
$N=1$ Supersymmetric Yang-Mills Theory,'' Phys. Lett. {\bf 113B} (1982) 231.}
that applies both to the two-dimensional ${\bf CP}^n$ model and to super 
Yang-Mills
domain walls.  \nref\becker{K. Becker, M. Becker, and A. Strominger,
``Five-Branes, Membranes, And Nonperturbative String Theory,''
Nucl. Phys. {\bf B456} (1995) 130, hep-th/9507158.}
\nref\morrison{K.  Becker, M. Becker, D. R. Morrison, H. Ooguri, Y. Oz, and Z. 
Yin,
``Supersymmetric Cycles In Exceptional Holonomy Manifolds And Calabi-Yau
Fourfolds,'' Nucl. Phys. {\bf B480} (1996) 225.} In section 4, we
 describe how a BPS-saturated domain wall could be represented
in the present formulation via a ``supersymmetric three-cycle'' in the sense
of \refs{\becker,\morrison}.  (But we do not actually prove the
existence of such a cycle obeying the appropriate boundary conditions.) 
The model  thus has both
one-branes -- the QCD strings -- and two-branes -- the domain walls.  
We establish by a topological argument a new result
which has not been guessed previously: the two-branes are
$D$-branes for the QCD string, that is a QCD string can end on a domain wall.  
This
is related to the behavior of the chiral order parameters inside the domain 
wall.
An intuitive explanation of this new effect
has been given by S.-J. Rey \ref\rey{S.-J. Rey, to appear.}
and will be mentioned in section 4.
Finally, we show that -- as one might have guessed intuitively -- the chiral
symmetry breaking of the model is reflected in a non-zero value of the
gluino condensate.  From a field theory point of view this is of course
an old result \ref\shifman{M. A. Shifman and A. I. Vainshtein,
``On Gluino Condensation In Supersymmetric Gauge Theories, SU(N) and O(N)
Groups,'' Nucl. Phys. {\bf B296} (1988) 445.}. 
To establish it via branes, we introduce
and evaluate an expression for the superpotential of an $N=1$ brane
configuration; this expression is likely to have other applications.

In section 5, it comes time to pay the piper.  Why has it been possible to
understand by semi-classical methods rather subtle, strong coupling properties 
of
QCD (or a good analog of it) that are traditionally  out of reach of such 
methods?
Are branes magic?  The answer is that what we have realized via
the branes is a generalization of the supersymmetric Yang-Mills theory that
depends on one extra parameter, essentially the Type IIA string coupling 
constant.
The limit in which the semi-classical methods of this paper are effective is
quite different from the limit in which the theory reduces 
to standard $N=1$ super Yang-Mills theory.  They both are (very likely) in the
same universality class, justifying the application of the results of sections
2-4 to the standard field theory, but quantitative details of the mass spectrum
and interactions will be different.  

The ``new'' theory is best understood in terms of compactification of the
six-dimensional $(0,2)$ supersymmetric field theory \ref\uwitten{E. Witten,
``Some Comments On String Dynamics,'' in the proceedings of {\it Strings '95},
ed. I. Bars et al., hep-th/9507121.} which can be interpreted in
terms of parallel fivebranes \ref\strom{A. Strominger,
``Open $p$ Branes,''  Phys.Lett. {\bf B383} (1996) 44-47, hep-th/9512059.}.  
The best approach
to the large $n$ limit of QCD might involve tackling first the large $n$ limit
of this theory.  It is also  at least remotely conceivable that
the sort of generalization of standard gauge theory considered
here could show up directly
in accelerator experiments (presumably with a small value of the deformation 
parameter and not the large value that makes possible semiclassical analysis
via branes), maybe even permitting new approaches to the gauge 
hierarchy problem.

The  brane approach to gauge theories is
not necessarily limited to supersymmetric theories.  To illustrate
this, we will in section 6 construct an $M$-theory brane
configuration relevant to  ordinary bosonic $SU(n)$ gauge theory
just as the configuration of section 2 is related to the supersymmetric
case.  As we explain, the mystery is not how to construct non-supersymmetric
brane configurations associated with interesting field theories, 
but what can be learned by studying them.

While carrying out the present work, I learned of  work by
K. Hori, H. Ooguri, and Y. Oz in which related results were obtained,
including a description 
of the brane configuration of section 2 and generalizations
to include chiral superfields \ref\hoo{K. Hori, H. Ooguri, and Y. Oz,
``Strong Coupling Dynamics of Four-Dimensional 
$N=1$ Gauge Theory from M Theory Fivebrane,'' hep-th/9706082.}.
Another closely related paper is 
\ref\biksy{A. Brandhuber, N. Itzhaki, V. Kaplunovsky, J. Sonnenschein,
and S. Yankielowicz, ``Comments On The $M$ Theory Approach to $N=1$ SQCD And
Brane Dynamics,'' hep-th/9706127.}.

\newsec{The Brane Configuration}
 
\subsec{Preliminaries}

We work in Type IIA superstring theory
with spacetime coordinates $x^0,x^1,\dots, x^9$.  
The brane configuration we will study is similar to ones discussed
in papers cited in the introduction.  In fact, it is a special
case (with some  branes omitted) of a configuration studied in \elitzur.   

We consider an NS fivebrane
located at $x^6=x^7=x^8=x^9=0$ (its worldvolume is thus spanned by
$x^0,x^1,\dots, x^5$) and 
an NS$'$ fivebrane located at $x^6=S_0$, $x^4=x^5=x^9=0$
(its worldvolume is thus spanned by $x^0,\dots ,x^3$ and $x^7,x^8$).
Here $S_0$ is an arbitrary length.  We set $v=x^4+ix^5$, $w=x^7+ix^8$.

Between the two fivebranes we suspend
$n$ Dirichlet fourbranes, whose worldvolumes are defined at the classical level
by $v=w=x^9=0$, $0\leq x^6\leq S_0$ (and so are spanned by 
$x^0,\dots,x^3$ along with $x^6$).  Quantum mechanically, $v,w,$ and $x^9$
are free to fluctuate as fields on the fourbrane, with boundary conditions
which, if one ignores fluctations in the fivebrane position, are 
$w=x^9=0$ at $x^6=0$ and $v=x^9=0$ at $ x^6=S_0$.

Interest will focus on the quantum field theory on the fourbrane.
This will be a four-dimensional field theory, since the fourbrane worldvolume
(being finite in the $x^6$ direction) is $3+1$-dimensional macroscopically.
Because of the $n$ parallel fourbranes which classically are coincident,
the theory will be an $SU(n)$ gauge theory.\foot{The
gauge group is really $SU(n)$ and not $U(n)$, as one might have supposed.  
This was explained at the $N=2$ level in \morewitten\ and continues to hold
after rotating the branes to get to $N=1$ supersymmetry.  
(Indeed, the $M$-theory brane configuration
found below has genus zero -- no massless photons -- and not genus 
one -- one massless photon.)  Apart from
arguments given in \morewitten, this can apparently be understood in terms of
Higgsing of the center of $U(n)$ by the
antisymmetric tensor fields on the fivebranes.}  The effective four-dimensional
theory has $N=1$ supersymmetry and has
no massless bosons other than the gauge fields
since all massless scalars on the fourbrane are projected out at $x^6=0$
($w$ and $x^9$) or at $x^6=S_0$ ($v$ and $x^9$).  So this theory is the
$SU(n)$ gauge theory with $N=1$ supersymmetry and no chiral multiplets.

The gauge coupling $g$ in this theory is related to the Type IIA gauge coupling
$g_{st}$ by
\eqn\omigo{{1\over g^2}={S_0\over g_{st}}}
up to a universal constant multiple.    
Since $g$ (which ultimately sets the QCD mass scale) is the only
parameter in the pure gauge theory, we see the fundamental fact that the
gauge theory depends on only one combination of $S_0$ and $g_{st}$.  
What happens in the brane theory will be examined in section 5.

The justification for claiming that this brane configuration can realize
the pure four-dimensional supersymmetric gauge theory -- decoupled from all
other complications of the string theory -- is as follows.  If $g_{st}$ is 
small
and $S_0$ is large, then the theory on the four-brane is weakly coupled
at the string   scale.  But, being infrared unstable, it flows at very low
energies to strongly coupled supersymmetric QCD.  At such low energies,
gravitation and all the other complications of the string theory can be
neglected.

\subsec{Solution Via $M$-Theory}

As in \morewitten, the solution of the model is made, at least at a formal
level, by going to $M$-theory.  We take $g_{st}$ large, whereupon
an eleventh dimension appears, a circle ${\bf S}^1$ with angular
 coordinate $x^{10}$.  We henceforth
measure distances in $M$-theory units.  We take
$0\leq x^{10}\leq 2\pi$ and put the radius $R$ of the circle in the
eleven-dimensional metric:
\eqn\bofo{ ds^2=\sum_{i,j=0}^9\eta_{ij}dx^idx^j+R^2(dx^{10})^2.}
If $C=2\pi R$ is the circumference of the circle, and $S$ is the separation
between the fivebranes measured in $M$-theory units, then the formula
\omigo\ becomes
\eqn\tomigo{{4\pi \over g^2}={S\over C}.}
(This assertion is a consequence of the assertion \ref\verlinde{E. Verlinde,
``Global Aspects Of Electric-Magnetic Duality,'' hep-th/9506011.}
that in compactification of a chiral two-form theory and hence of a fivebrane 
on
a two-torus, the induced $\tau$ parameter of the four-dimensional $U(1)$ 
gauge field derived from the two-form is the $\tau$ of the two-torus.  In the
present case, the fourbrane corresponds to a long tube connecting the 
fivebranes,
of circumference and length $C$ and $S$, and this gives \tomigo.) 

Going to $M$-theory makes possible a solution of the theory as follows.
In $M$-theory the fourbranes are just fivebranes wrapped
on the ${\bf S}^1$.  In fact, all branes in the problem are fivebranes,
and the fivebrane world-volume can be described as ${\bf R}^4\times \Sigma$
where ${\bf R}^4$ is a copy of four-dimensional Minkowski space
(with the coordinates $x^0,\dots,x^3$ that the various branes have in common)
and $\Sigma$ a two-dimensional surface.  $\Sigma$ is in fact a complex
Riemann surface in a three-dimensional space with coordinates $v,w$,
and $t=e^{-s}$, where $s=R^{-1}x^6+ix^{10}$.\foot{The $R^{-1}$ is present
in this formula because of the $R^2$ multiplying $(dx^{10})^2$ in \bofo.}

To determine $\Sigma$, we may simply proceed as follows.  On $\Sigma$,
$v$ goes to infinity only at one point, which is infinity on the original
NS fivebrane; and $v$ has only a single pole there, since there is only
one NS fivebrane.  So $v$ is a holomorphic function on $\Sigma$ with only
one pole at infinity.  $\Sigma$ can therefore be identified as the complex
$v$-plane (perhaps with some points deleted where other variables have
poles).  Likewise $w$ has only a single pole -- as $w\to \infty$ only on the
NS$'$ fivebrane.  $v=0$ where $w$ has a pole (since the NS$'$ fivebrane is
at $v=0$) and likewise $w=0$ at $v=\infty$.  The most general
function with these properties is $w=\zeta v^{-1}$ for some complex
constant $\zeta$.
$\Sigma$ should go to infinity only at the positions of one of the two
fivebranes, so $\Sigma$ is precisely the $v$-plane with the two points
$v=0$ and $v=\infty$ deleted.

$t$ should be given by a formula $t=F_n(v)$, where $F_n(v)$ is such that
the equation $F_n(v)=t$, regarded as an equation for $v$ with fixed $t$,
has $n$ roots -- corresponding in the Type IIA description 
to the fact that for given $t$, there
are $n$ fourbranes.  Any $F_n$ obeying this condition is of the form
$F_n(v) =P(v) /Q(v)$, where $P$ and $Q$ are polynomials of degree $n$.
Since $t=\infty$ and $t=0$ are at infinity (they
correspond to $x^6=\pm\infty$), $t$ should have no zeroes or poles
as a function of $v$ except at $v=0$ (which is $w=\infty$) or $v=\infty$.  The 
most
general possibility obeying these conditions is essentially
$t=v^n$.  (A multiplicative constant, that is $t=cv^n$, can be absorbed
in adding a constant to $x^6$; this acts by $t\to \lambda t$. The alternative
solution $t=v^{-n}$ is equivalent modulo an exchange of the two fivebranes.
The solution $t=v^n$ corresponds to having the NS fivebrane at smaller
$x^6$ than the NS$'$ fivebrane, so that $x^6\to -\infty$ and $t\to \infty$
 on the NS fivebrane while $x^6\to \infty$ and  $t\to 0$ on the NS$'$ 
fivebrane.)

The curve $\Sigma$ is thus described by the equations
\eqn\humbo{ \eqalign{ w & =\zeta v^{-1}\cr
                      v^n & = t\cr
                      w^n & =\zeta^n t^{-1}.\cr}}
Of course, this description is redundant; the second or third equation
could be dropped.

\bigskip\noindent{\it Symmetries}

Now, let us analyze the behavior at infinity.  Infinity has two components,
$v\to\infty$ and $w\to\infty$.  One component of infinity can be described
asymptotically by $w\to 0$ and $v^n=t$.  The other component at infinity
can be described asymptotically by $v\to 0$ and $w^n=\zeta^n t^{-1}$.
We see that only $\zeta^n$, and not $\zeta$, enters in the behavior at
infinity.

In studying quantum field theory on branes, the behavior at infinity
determines a particular quantum problem.
After fixing the behavior at infinity, one looks for the possible
supersymmetric or lowest area branes with the given asymptotic behavior.
They represent possible quantum ground states in the quantum problem in 
question.  

In the case at hand, for given behavior at infinity, there are $n$ possible
choices for the interior behavior of the brane.  Indeed, the number
$\zeta^n$ appears in the behavior at infinity.  The possible interior
behaviors or quantum states correspond to the $n$ possible values of $\zeta$
for given $\zeta^n$.  

The supersymmetric $SU(n)$ gauge theory that we are trying to solve is
believed to have $n$ vacua resulting from a spontaneously broken
${\bf Z}_n$ chiral symmetry.  It is natural to try to identify the
$n$ vacua that we have just found with the $n$ vacua expected from
chiral symmetry breaking.

The symmetries of the quantum problem are symmetries of the
asymptotic behavior of the branes at infinity.  A symmetry is unbroken
if it leaves fixed the entire brane, and not just the behavior at infinity.
In our case, there is a $U(1)$ symmetry group $U$ that acts
by
\eqn\hungo{\eqalign{t\to &e^{in\delta} t\cr
                    v\to &e^{i\delta} v \cr
                    w\to &e^{-i\delta}w.  \cr}}
There is also an additional ${\bf Z}_n$ symmetry group $H$ generated by
\eqn\pungo{\eqalign{t\to &t \cr
                    v\to &v \cr
                    w\to &\exp(2\pi i/n)w.\cr}}
Finally there is a ${\bf Z}_2$ symmetry  $K$ that acts by
\eqn\onungo{\eqalign{v\to & w \cr
                     w\to & v \cr
                     t\to & \zeta^nt^{-1}.\cr}}
Of these symmetries, the ${\bf Z}_n$ is spontaneously broken; it 
is a symmetry of the brane at infinity but is not an exact symmetry of the
brane as it does not leave invariant the first equation in \humbo.
The other symmetries are unbroken, as they are invariances of \humbo.

To identify these symmetries, first note the following.  Let $\W$ be the
complex three-fold with coordinates $v,w,t$. Note that topologically
$\W=\R^5\times \S^1$.   $\W$  can be regarded as a 
flat Calabi-Yau manifold with metric
\eqn\impo{ds^2=|dv|^2+|dw|^2+R^2|ds|^2 =|dv|^2+|dw|^2+R^2{|dt|^2\over |t|^2}}
and holomorphic three-form
\eqn\unimpo{\Omega=R{dv\wedge dw\wedge dt\over t}.}
(Thus, $\Omega\wedge\overline\Omega$ is the volume form of the Riemannian
metric on $\W$.)  As in Calabi-Yau compactification of superstring theory,
the superspace measure $d^2\theta$ (the $\theta$'s being the chiral odd
coordinates of $N=1$ superspace) transforms like $\Omega^{-1}$, and a symmetry
is an $R$-symmetry if and only if it transforms $\Omega$ non-trivially.
By this criterion, we see that the spontaneously broken ${\bf Z}_n$
symmetry group found above is a group of $R$-symmetries;
this is the expected spontaneously broken ${\bf Z}_n$ chiral symmetry
group of the model.\foot{The symmetry group of the theory is actually
${\bf Z}_{2n}$, but the element $-1$ of ${\bf Z}_{2n}$ is a $2\pi$ rotation
of spacetime and is unbroken.  The group that acts on bosonic variables
such as the gluino bilinear $\Tr \lambda\lambda$
is the quotient $H={\bf Z}_n$ of ${\bf Z}_{2n}$, and this from a field
theory point of view is the spontaneously broken chiral symmetry group.}

What about the other symmetries?  $K$, since it reverses
the orientation of the Type IIA spacetime (whose coordinates are the
original ten variables $x^0,\dots,x^9$), 
reverses the orientation of all elementary
Type IIA strings.  Gauge bosons and gluinos in the adjoint representation
of $SU(n)$ are built from elementary strings that connect different
fourbranes.  Reversing the orientation of the elementary strings
exchanges positive and negative roots of $SU(n)$ and hence
exchanges the fundamental and antifundamental
representations of $SU(n)$.  This operation is  usually
called ``charge conjugation.''  It is expected to be unbroken in the $SU(n)$
gauge theory, and this agrees with what we have just found.

More mysterious at first sight is the proper interpretation of the $U(1)$.
In the Type IIA description, there appear to be three $U(1)$ symmetries.
Rotations of the $v$-plane, the $w$-plane, and the $x^{10}$ circle
appear to be separate invariances of the classical brane configuration.
In actuality, one does not have all three $U(1)$ symmetries separately
because the end of a fourbrane on a fivebrane looks like a ``vortex.''
In fact, on the two fivebranes one has respectively $x^{10}\sim n \,\,{\rm arg}
(v)$ for large $v$ and 
$x^{10} \sim -n \,\,{\rm arg}\,\ln(w)$ for large $w$.
These expressions are invariant only under one linear combination of
the three $U(1)$'s, namely the one defined in \hungo.  

This one $U(1)$ symmetry  acts trivially on both
gluons and gluinos and hence on the entire $SU(n)$ gauge theory that we
are aiming to investigate in the present paper.  This is clear
in the Type IIA description.  States of the Type IIA
superstring theory on which this symmetry acts
non-trivially are, for instance, zerobranes that carry momentum in the
$x^{10}$ direction, and all sorts of things that carry angular momentum
in the $v$-plane or $w$-plane.  Gluons and gluinos are invariant under this 
symmetry.
In any limit in which the brane theory
actually reproduces the supersymmetric Yang-Mills theory quantitatively, 
all modes carrying the $U(1)$ charge must decouple from the Yang-Mills physics.

\subsec{Rotation Of $N=2$ Solution}

For completeness, we will now show how the brane configuration that
we found directly could have been found by ``rotating,'' in the sense of
\ref\rotat{J. L. F. Barbon, ``Rotated Branes And $N=1$ Duality,''
hep-th/9703051.}, 
the brane configuration that describes $SU(n)$ gauge theory with
$N=2$ supersymmetry.  We recall \morewitten\ that the latter is described
by the brane configuration
\eqn\hobo{t^2+P_n(v)t+1 = 0}
in $v-t$ space, at $w=0$. This describes $n$ fourbranes suspended between
two parallel fivebranes.  $P_n(v)$ is a polynomial of the form
\eqn\ugg{P_n(v)=v^n +u_2v^{n-2}+\dots +u_n,} where the $u_i$ are the ``order
parameters'' of the theory.  

 ``Rotating'' one of the fivebranes will break
$N=2$ supersymmetry to $N=1$.  This can only be done at points in moduli
space at which the genus $n-1$ curve \hobo\ degenerates to a curve of
genus zero \ref\sw{N. Seiberg and E. Witten,
``Electric-Magnetic Duality, Monopole Condensation, 
And Confinement In $N=2$ Supersymmetric Yang-Mills Theory,''   
Nucl. Phys. {\bf B426} (1994) 19. }.  A curve $\Sigma$ of genus
zero can be ``parametrized rationally.'' 
This means the following.  One can introduce
an auxiliary complex parameter $\lambda$, and identify $\Sigma$ as the
complex $\lambda$ plane, perhaps with some points deleted or
a point at infinity added.  Then $v,w,$ and $t$ can be expressed as rational
functions of $\lambda$.  

As $v$ has poles only at ``ends'' of fivebranes, of which there are precisely
two, $v$ has exactly two simple poles as a function of $\lambda$.
We can hence take $v=\lambda+c\lambda^{-1}$, for some constant $c$,
 with no loss of generality.
Also, $t$ can go to zero or infinity only at poles of $v$, that is
at $\lambda=0$ or $\infty$; this implies
that $t$ is a constant multiple of a power of $\lambda$.
Requiring that \hobo\ should be obeyed for $\lambda\to\infty$ and $\lambda\to 
0$
(and recalling the form \ugg\ of $P_n$),
we find that $t=-\lambda^n$ and that $c^n=1$, and moreover for each choice
of $c$ the polynomial $P_n$ is uniquely determined.  (There are also
solutions with $t$ a multiple of $\lambda^{-n}$ but these differ
by $\lambda\to \lambda^{-1}$ and merely give another way to parametrize
the same branes.)

Thus, as expected we have found $n$ polynomials $P_n$, corresponding
to the $n $ roots of $c^n=1$, for which the curve $\Sigma$ degenerates to genus
zero and hence can be ``rotated.''  We now set $c=1$, for brevity, and
attempt to make the rotation.

After the rotation,  $w$, instead of being zero, should
be a non-zero holomorphic function on $\Sigma$.  As we only
want to rotate one fivebrane, $w$ should get a pole only at one ``end''
of $\Sigma$ and should vanish at the other end.  There is hence no
essential loss of generality in setting $w=\zeta \lambda^{-1}$ for
some complex constant $\zeta.$  Moreover, after redefining
$v$ by $v\to v-\zeta^{-1} w$, we reduce to $v=\lambda$, so finally (after a 
sign change of $t$)
the equations defining the curve are the ones found above:
$w=\zeta v^{-1}$, $t=v^n$.

\subsec{Mass Scale}

In the forgoing, we have simply taken $\zeta$ to be an arbitrary complex 
number.
As long as one considers only the universality class of the model, the
value of $\zeta$ does not matter.  If one considers only holomorphic
quantities, $\zeta$ can be scaled out by redefining $w$.  

Such a scaling will, however,
affect the particle masses and interactions,
as it does not leave invariant the metric of space-time.
In fact, the parameters of the model from the $M$-theory point of view are
$\zeta $ as well as $R$, the radius of the eleventh dimension.  We cannot
expect to get a quantitative
description of supersymmetric Yang-Mills theory if these are taken
to be generic (say of order one in $M$-theory units), since that will give
no mechanism to decouple all of the extra degrees of freedom and complications 
of
$M$-theory.  One certainly can recover super Yang-Mills theory for
$R\to 0$ (weakly coupled Type IIA superstring theory) and small $\zeta$.
More details about the extent to which this theory
behaves like super Yang-Mills theory will gradually become clear.

\newsec{Confinement}

\subsec{\it Membranes And Strings}

One of the main mysteries of QCD is confinement. Accordingly, we would
now like to analyze confinement in the present context.
%%%%

The signal of confinement is that the theory should contain strings, which
we will call QCD strings\foot{From the discussion below and
in section 6, it appears 
likely that the strings we consider are in the same universality class as the
conventional non-supersymmetric
QCD string, at least if $R^2\zeta^{1/2}$ is sufficiently small.}, 
which could terminate on an external quark (a charge in the 
fundamental representation of $SU(n)$),
but which, in the absence of dynamical quarks, are stable and cannot break.
Thus, an external quark and antiquark connected by such a string and
separated a distance $r$ would have an energy linear in $r$, a phenomenon
referred to as confinement.

Since $n$ copies of the fundamental representation of $SU(n)$ can combine
to a singlet, the QCD string is only conserved modulo $n$; $n$ parallel
QCD strings can annihilate.

\def\R{{\bf R}}

\nref\town{P. Townsend, ``$D$-Branes From $M$-Branes,'' Phys. Lett.
{\bf B373} (1996) 68, hep-th/9512062.}
A natural guess, in the present context, is that the QCD string should
be identified as an $M$-theory membrane, whose boundary is on the fivebrane
as in \refs{\strom,\town}.  
%%%%%

The membrane lives in our $M$-theory world
$\R^4\times \W\times \R$, where $\R^4$ (parametrized by $x^0,\dots,x^3$) is the
effective Minkowski space, $\W$ is a complex three-fold, with coordinates
$v,w$, and $t=e^{-s}$, that contains the complex curve $\Sigma$, and the last
factor $\R$ (parametrized by the last coordinate $x^9$) plays no role in the 
present paper.    

We consider a membrane that is the product of a string or onebrane in $\R^4$
times a onebrane in $\W$.  Such a membrane will look like a string to a
four-dimensional observer.  We could and eventually will consider
closed onebranes in $\W$.
But more immediately
pertinent to the QCD problem are
open one-branes that end on $\Sigma$.  We consider
an open curve $C$ in $\W$, parametrized by a parameter $\sigma$ 
with $0\leq \sigma\leq 1$; we orient $C$
 in the direction of increasing $\sigma$. 
We require that the endpoints of $C$ (the points with $\sigma=0,1$) are
 in $\Sigma$.
With $\Sigma$ defined by the familiar equations
\eqn\unu{\eqalign{w & = \zeta v^{-1} \cr
                  v^n & = t \cr
                  w^n & = \zeta^nt^{-1},\cr}}
we consider a curve $C$ determined by setting $t$ to a fixed  constant
$t_0$ ($t_0$ is necessarily non-zero,  since
the point $t=0$ is at $w=\infty$ and so is omitted), and
\eqn\jubu{\eqalign{ v&=t_0^{1/n}\exp(2\pi i \sigma/n)\cr
      w& =\zeta v^{-1}.\cr}}
Here $t_0^{1/n}$ is any fixed $n^{th}$ root of the  complex nuber
$t_0$.

%%%%%%
By picking this particular $C$, we get a string in spacetime which we will
call the candidate QCD string.  We will
show that $n$ such strings can annihilate and then show that they can 
annihilate
only in groups of $n$.\foot{Twisted open strings that can annihilate
in groups of $n$ were described in another context in 
\ref\blum{J. Blum,   ``$F$ Theory Orientifolds, $M$ Theory Orientifolds,
And Twisted Strings,'' hep-th/9608053.}.}

Consider $n$ strings determined by curves $C_j\subset \W$, $j=1, \dots,n$
with the $C_j$ defined by $t=t_0$ and
\eqn\hubu{\eqalign{v=&t_0^{1/n}\exp(2\pi i\sigma/n)\exp(2\pi i j/n)\cr
         w& =\zeta v^{-1}. \cr}}
The $C_j$'s all have boundary on $\Sigma$, so they all determine strings
in $\R^4$.  The boundaries of the $C_j$ (with boundary points weighted
by orientation) add up to zero, so these $n$ strings can detach themselves
from $\Sigma$, forming a closed loop in the $v $-plane.  This loop can be
contracted to a point, so the $n$ strings determined by the $C_j$ can 
annihilate.

%%&&&
On the other hand, the $C_j$ are all homotopic to each other (by a homotopy
that keeps the boundary on $\Sigma$).  A homotopy between them can be made by
taking $t_0\to t_0e^{2\pi i\alpha}$, with $0\leq\alpha\leq 1$.  This homotopy
deforms $C_j$ to $C_{j+1}$.
So the process described in the last paragraph represents the annihilition
of $n$ identical candidate QCD strings.

%%&&&&
\def\Z{{\bf Z}}
To show that these candidate QCD strings
can annihilate {\it only} in groups of $n$, I will introduce some mathematical
machinery that will have further applications below.  Given a topological
space $\W$, one defines the homology groups $H_k(\W,\Z)$
to consist of closed $k$-dimensional submanifolds (or more generally 
$k$-cycles)
in $\W$ modulo boundaries of $k+1$-manifolds.  Homology groups 
$H_k(\Sigma,\Z)$ of
a subspace $\Sigma$ of $\W$ are defined in the same way, using cycles in 
$\Sigma$.
Finally, one defines the relative homology groups $H_k(\W/\Sigma,\Z)$ to 
consist
of $k$-dimensional submanifolds  of  $\W$ (or more generally chains)
which are not necessarily closed, but whose
boundaries are in $\Sigma$ (modulo an equivalence relation in which
a chain or submanifold
is considered trivial if after adding a chain that lies in $\Sigma$
it is a boundary).  These homology groups are linked by an exact sequence
\eqn\dofo{\dots \to H_{k+1}(\W/\Sigma,\Z)\to H_k(\Sigma,\Z) 
\to H_k(\W,\Z)\to H_k(\W/\Sigma,\Z)
\to H_{k-1}(\Sigma,\Z) \to \dots .}
The map $H_k(\Sigma,\Z)\to H_k(\W,\Z)$  maps a cycle
in $\Sigma$ to the ``same'' cycle in $\W$; the map $H_k(\W,\Z)\to 
H_k(\W/\Sigma,\Z)$
maps a cycle in $\W$ (which has no boundary) to the ``same'' cycle,
regarded as a relative cycle in
$\W/\Sigma$ (i.e., as a submanifold or chain which
is allowed to have a boundary in $\Sigma$ but may have zero
boundary);
finally the map from $H_k(\W/\Sigma,\Z)$ to $H_{k-1}(\Sigma,\Z)$ maps a 
$k$-dimensional submanifold $S$ of $\W$ whose boundary 
is in $\Sigma$ to the boundary
of $S$, which is a $k-1$-dimensional closed manifold in $\Sigma$.

In $M$-theory on $\R^1\times X$, where $\R^1$ is ``time'' and $X$ is
a ten-manifold  representing ``space,'' membranes are classified topologically
by $H_2(X,\Z)$.\foot{If $X$ is unorientable, one wants homology with twisted
coefficients, since the three-form field of eleven-dimensional supergravity
(which couples to the membrane world-volume) is odd under parity.  In the 
present
paper, spacetime will always be orientable.}
If in space-time there is a fivebrane with world-volume $\R^1\times B$,
so that the membranes can have boundary on $B$, then membranes are classified
topologically by $H_2(X/B,\Z)$.  In our case, 
$X=\R^4\times \W$ and $B=\R^4\times \Sigma$.  Moreover,
we want a membrane that is a product
of a string in $\R^4$ times a onebrane in $\W$. 
These are classified topologically
by $H_1(\W/\Sigma,\Z)$.  

\def\S{{\bf S}}
To compute this group, we use the exact sequence \dofo.  The relevant
part of this exact sequence reads
\eqn\pollo{\dots  \to H_1(\Sigma,\Z)\to H_1(\W,\Z)\to H_1(\W/\Sigma,\Z)\to 0.}
(We are here using the fact that as $\Sigma$ is connected, 
the map from $H_0(\Sigma,\Z)$ to $H_0(\W,\Z)$ is an isomorphism, both groups
being isomorphic to $\Z$; so the map from $H_1(\W/\Sigma,\Z)$ to
$H_0(\Sigma,\Z)$ is zero.)  Hence if $i:H_1(\Sigma,\Z)\to H_1(\W,\Z)$ is the 
map
in \pollo, then 
\eqn\onollo{H_1(\W/\Sigma,\Z) \cong H_1(\W,\Z)/i(H_1(\Sigma,\Z)).}
Now, with $\Sigma$ being the complex $v$ plane with the origin deleted,
$H_1(\Sigma,\Z)$ is isomorphic to $\Z$, being generated by a simple closed
curve that wraps once around the origin in $v$.  With $\W=\R^5\times \S^1$,
$H_1(\W,\Z)$ is likewise isomorphic to $\Z$, being generated by a simple
closed curve that wraps once around the $\S^1$ factor at a fixed point in
$\R^5$.  But  the angular
variable on the $\S^1$ is the phase of the complex variable $t$.  So the 
formula
$t=v^n$ means that the map $i$ multiplies a loop that wraps once around the
origin in the $v$-plane to a loop that wraps $n$ times around the origin in the
$t$-plane.  In other words, $i$ is multiplication by $n$ and hence
\eqn\onollo{H_1(\W/\Sigma,\Z)\cong \Z_n.}

The argument also shows that the generator of $H_1(\W/\Sigma,\Z)$ is the image
in that group  of a generator of $H_1(\W,\Z)$, that
is of a loop
$C'$ that wraps once around the $\S^1$.  Such a $C'$ can take the form
\eqn\ilmop{\eqalign{ v& =t_0^{1/n} \cr  
                     w & =\zeta v^{-1} \cr 
                     t& = t_0\exp(-2\pi i\sigma).\cr}}
This is actually a closed one-brane, the boundary values at $\sigma=0$ and $1$
being equal; 
it merely corresponds to a closed membrane wrapping once around the eleventh
dimension, and  is identified
in double dimensional reduction \ref\double{M. J. Duff, P. S. 
Howe, T. Inami and K. S. Stelle, ``Superstrings in $D=10$ from 
supermembranes in D=11,'' Phys. Lett. {\bf B191} (1987) 70.}  with the
elementary Type IIA  string! 
To make sense of this, first note that the curve $C'$ really is homotopic
(via a family of open curves with boundary on $\Sigma$) to $C$.  Such
a homotopy can be made by the one-parameter family of curves $C_\alpha$,
$0\leq\alpha\leq 1$, with 
\eqn\nimlop{\eqalign{v & = t_0^{1/n}\exp(2\pi i \alpha\sigma/n)\cr
                     w& =\zeta v^{-1}\cr  
                     t& = t_0 \exp(2\pi i (\alpha-1)\sigma).\cr}}
Here $C_0=C'$ and $C_1=C$.  So it is true that the elementary Type IIA
superstring is equivalent topologically to the candidate QCD string.  What
does this mean?

\bigskip\noindent{\it Energy Scales}

To proceed farther, we really need a discussion of energy scales in this
problem.  

The loop $C'$ corresponds to a string in spacetime with a tension $T'$ 
 which (in $M$-theory units) is of order 
\eqn\huj{T'\sim 2\pi R}
with $R$ the radius of the eleventh dimension.

What about the tension of a string built using the curve $C$?  This is 
proportional
to the length of $C$.  In traversing $C$,  $t$ is constant, $v$ changes
by an amount of order $t_0^{1/n}/n$, and $w=\zeta v^{-1}$ changes by
an amount of order $\zeta t_0^{-1/n}/n$.  To minimize the length of $C$, we 
must
pick $t_0$ of order $\zeta^{n/2}$, whereupon the length of 
$C$ is of order $\zeta^{1/2}/n$.  The tension $T$ of a four-dimensional string
built using $C$ is hence of order
\eqn\jiko{T\sim {|\zeta|^{1/2}\over n}.} 
Actually, this formula is only valid in a regime in which the string
can be treated semiclassically, which as we discuss in section 5
will not  be the case for all values of the parameters.  
(As we discuss momentarily, the candidate
QCD string is not BPS-saturated, so its tension is subject to renormalization.)
We will actually not determine the precise conditions for validity of \jiko.

To get a reasonable quantitative likeness of QCD, we need $T<<T'$,  
so that a string built from $C$ (which is localized near the fivebrane)
is energetically favored over a string built from $C'$ (which is an elementary
Type IIA string, free to wander in ten dimensions and having nothing much to
do with QCD).  Thus, while the parameters can possibly vary more widely
without spoiling the ``universality class'' of the theory, to get a 
quantitative
likeness of QCD, we must require at least that
\eqn\ogiko{{|\zeta|^{1/2}R\over n}<<1.}

If this inequality holds, then the candidate QCD string is much lighter than 
the
elementary Type IIA string, and there is at least some hope that the QCD sector
of the theory is decoupled from the complications of superstring theory and
$M$-theory.  In this situation, the candidate QCD string is, from this point 
of view,
a sort of bound state of an elementary Type IIA string with a fivebrane,
in which the binding energy is almost one hundred percent
 (of the energy of a Type IIA
string far from the fivebrane).  There is the further curious fact that 
although
in the absence of the fivebrane, the number of long parallel
elementary Type IIA strings
would be conserved, in the presence of the fivebrane they can turn into
QCD strings and annihilate in groups of $n$.

\subsec{Universality Class Of The QCD String}

Now we would like to look more critically at the worldsheet properties
of our candidate QCD
strings and compare to what is expected for actual supersymmetric QCD strings.

Although it is possible to have BPS-saturated strings in $N=1$ supersymmetric
models in four dimensions, the supersymmetric Yang-Mills theory without
chiral superfields has no conserved charges that could appear as central
charges for strings.  Also, the fact that the QCD string is only conserved 
modulo
$n$ shows that it does not carry an additive conserved quantity and so
cannot have a central charge.  The supersymmetric QCD string of the minimal
$N=1$ super Yang-Mills theory is therefore
not invariant under any of the four supersymmetries.  So on this string
propagate a full set of Goldstone fermions (two left-moving ones and two 
right-moving
ones).  It is not clear whether there are additional massless modes on the 
supersymmetric QCD string.

What about the candidate QCD string?  At first sight,
it appears to have one crucial difference from the supersymmetric QCD string.
As described in eqn. \hungo, the brane configuration has a $U(1)$ symmetry $U$
that  acts trivially on all QCD excitations.  
In particular, for a string
to be interpreted as a product of QCD only (independent of
all the other degrees of freedom of $M$-theory), it must be invariant
under $U$; and $U$ must act trivially
on the massless degrees of freedom that propagate on the string.

Since $U$  acts on
$t_0$ by $t_0\to e^{in\delta}t_0$, any given classical configuration of
our candidate QCD string is not invariant under $U$.
It appears that $U$ is spontaneously broken along the string, in which case
$U$ would certainly act non-trivially on one of the massless modes that
propagates on the string, namely the Goldstone boson.

To rescue the hypothesis that our candidate is really in the universality
class of the QCD string, we must show that the dynamics on the string
is such that at sufficiently long wavelengths all degrees of freedom that
propagate on the string and on which $U$ acts non-trivially get mass.
There is no issue of whether $U$ is spontaneously broken or not; by Coleman's
theorem, continuous symmetries such as $U$ are not spontaneously broken
in two spacetime dimensions, so the $U$ symmetry, even though spontaneously
broken classically, will be restored at sufficiently long distances.
The issue is whether {\it after} this symmetry restoration, all degrees
of freedom propagating on the string that couple to $U$ get mass (which
would agree with a conventional supersymmetric QCD string) or there remain
massless excitations on the string on which $U$ acts non-trivially (in which
case our candidate string is not in the universality class of the conventional
supersymmetric QCD string).  

The candidate QCD string appears to differ in yet a second way from actual
QCD strings.  It appears at first sight that the candidate QCD string has
a conserved winding number that measures the number of times that it wraps
around the circle factor in $Y=\R^5\times \S^1$.  This  has no
analog in QCD, and must somehow disappear to justify a claim that the candidate
QCD string has the universality class of an actual QCD string.  We will
see that this question is closely linked to the decoupling of $U$.

\bigskip\noindent{\it Decoupling The $U(1)$ And Generating A Mass Gap}

A model with a spontaneously broken $U(1)$ global symmetry at tree level
has a classical Lagrangian that looks like
\eqn\odo{L_2={r^2\over 4\pi}\int d^2x \,|d\psi|^2,}
where $\psi$ is an angular variable that is rotated by the $U(1)$,
and $r$ is a constant;
 there can be additional  terms of higher order or involving massive fields or
massless but $U(1)$-invariant fields.
Any such theory in $1+1$ dimensions has symmetry restoration quantum
mechanically. Under what conditions
does it develop a mass gap, as  a result of which the
$U(1)$ symmetry can decouple from the low energy physics?  

If one did not have the $U(1)$ symmetry, the obvious perturbation
of \odo\ that produces a mass gap would be a $\cos m\phi$ perturbation for some
integer $m$; this
is a relevant operator if $r$ is large enough, and its addition to the 
Lagrangian
produces a mass gap.  However, all such terms are forbidden by the $U(1)$
symmetry.  

There is a slightly less obvious type of perturbation of \odo\ that can induce
a mass gap.  This is the {\it dual} (under the usual $r\to 1/r$ two-dimensional
$T$-duality) of a $\cos m\phi$ perturbation. Duality exchanges $\cos m\phi$
with ``twist fields.''  If $r$ is sufficiently {\it small}, the twist
fields are relevant operators and can induce a mass gap.  While invariant 
under the $U(1)$, the twist fields do not preserve the winding number $\oint 
d\psi/
2\pi$.  The free theory \odo\ has no other potentially relevant perturbations.

The conclusion, then, is that in a model that classically has a spontaneously
broken $U(1)$, the $U(1)$ can decouple at low energies if and only if $r$ is
sufficiently small and the winding number is not conserved.

The importance of violation of the winding number conservation
can alternatively be explained as follows.  
(This explanation will perhaps make it obvious that  adding more
matter fields to \odo\ does not change the picture.) 
Let $|\Omega\rangle$ 
be the vacuum of a model with a $U(1)$ symmetry in two dimensions, 
and let $f$ be a  function
of the spatial coordinate $x$ that is $0$ at $x=-\infty$ and $1$ at 
$x=+\infty$.
Let $J$ be the conserved current that generates the $U(1)$ symmetry and $J_0$
the charge density.  Consider the state
\eqn\hoggo{|\Psi\rangle=\exp\left(2\pi i\int_{-\infty}^\infty dx \,f(x)J_0(x)
\right)|\Omega
\rangle.}
In the limit that $f$ is very slowly varying, the energy of this state 
converges
to zero (that is, to the energy of $|\Omega\rangle$).  The state $|\Psi\rangle$
carries winding number one, if the winding number is conserved.  So either
(i) the winding number is not conserved, or (ii) the current acting on the 
vacuum
can create excitations of arbitrarily low energy, and thus the $U(1)$ symmetry
does not decouple from the very low energy two-dimensional physics.

A necessary condition for the candidate QCD strings
 to be in the universality class of actual supersymmetric QCD strings is
therefore that there should be no conserved winding number on the string 
associated
with the $U(1)$ symmetry $U$.  
This is necessary for the decoupling of $U$, and (as we have already noted)
is desireable in itself as
this winding number has no analog in QCD.

We must show, then,  that a candidate QCD string
that wraps  around the $\S^1$ (in $\W=\R^5\times \S^1$) can
be unwrapped.  To investigate this, we let $\sigma$ and $\rho$ be the
two coordinates on an $M$-theory membrane.  We have so far suppressed $\rho$,
describing the membrane at fixed $\rho$ as a curve 
\eqn\hogo{C: ~ v=t_0^{1/n}e^{2\pi i\sigma/n},\,\,w=\zeta v^{-1},\,\,\,t=t_0.}
Now as $\rho$ varies, the curve will move in $\R^4$ (which is why a 
four-dimensional
observer interprets it as a string) and in addition $t_0$ can change.
By giving $t_0$ an appropriate $\rho$ dependence, we can describe
a membrane that represents a string that wraps around the circle:
\eqn\gogo{\eqalign{ v & = t_0^{1/n}e^{2\pi i(\sigma/n+\rho)} \cr
                    w & = \zeta v^{-1}\cr 
                    t & = t_0 e^{2\pi in\rho}.\cr}}
This string actually wraps $n$ times around the circle, because of the exponent
in the formula for $t$. 

To show that such a wrapped string can actually disappear, it suffices
to show that the two ends of the membrane loop around the same closed curve
in $\Sigma$, so that the membrane can detach itself from $\Sigma$ and turn
into a closed membrane in spacetime.  Having
done so, the membrane represents a class in $H_2(\W,\Z)$, which (for
$\W=\R^5\times \S^1$) vanishes, so the membrane can annihilate.

In fact, at $\sigma=0$, the membrane traverses the loop
$v=t_0^{1/n}e^{2\pi i\rho}, \,t=t_0e^{2\pi in\rho}$.  At the other end
of the membrane, $\sigma=1$, the membrane traverses the loop
$v=t_0^{1/n}e^{2\pi i(1/n+\rho)}, \,t=t_0e^{2\pi in\rho}$.  After
a transformation $\rho\to \rho+1/n$, these coincide, showing that the
membrane is executing the same loop at both ends and so can indeed detach
itself and annihilate.

Since the above discussion was somewhat {\it ad hoc} and appears to deal
only with the case that the winding number is a multiple of $n$, we will now
carry out the analysis using a more powerful mathematical language.
Winding states of the membrane are classified topologically by
$H_2(\W/\Sigma,\Z)$.  Since $H_2(\W,\Z)=0$, the exact sequence \dofo\ says
that this group is the kernel of the map $i:H_1(\Sigma,\Z)\to H_1(\W,\Z)$.
We have already computed that that map is multiplication by $n$, and in 
particular
is injective, so $H_2(\W/\Sigma,\Z)$ vanishes and the membrane has no conserved
winding numbers.

So there is no topological obstruction to the hypothesis that the $U$ symmetry
decouples completely from the very long wavelength physics on the membrane.
Indeed, it should be expected to do so if the effective value of $r$ is
sufficiently small.  The effective $r^2$ is actually the string tension
$T$ times $R^2$.  Thus the quantity that should be sufficiently small is
\eqn\gujo{y=TR^2={|\zeta|^{1/2}R^2\over n},}
a combination that we will meet again in section 5.

\newsec{Domain Walls}

\subsec{Preliminaries}

A consequence of having a spontaneously broken discrete symmetry is that
there can be domain walls separating spatial domains containing different 
vacua.
In this section, we will consider these  domain walls in super Yang-Mills 
theory.

If one contemplates these domain walls in the context of the $1/n$ expansion
of the $SU(n)$ theory, one immediately runs into a puzzle.
\foot{Somewhat analogous questions about counting of powers of $n$, together
with the analogy between string theory and the $1/n$ expansion,
 were part of the 
motivation for some early pre-$D$-brane work 
(see for example
\ref\green{M. B. Green, ``Space-Time Duality And Dirichlet String Theory,''
Phys. Lett. {\bf B266} (1991) 325, ``Point-Like States For Type II 
Superstrings,''
Phys.Lett. {\bf B329} (1994) 434. })
 and
for early conjectures \ref\shenker{S. H. Shenker, ``The Strength Of
Nonperturbative Effects In String Theory,'' in {\it Cargese 1990, Proceedings,
Random Surfaces And Quantum Gravity.}} about string theory
 corrections of order $e^{-1/g}$.} 
  It is believed
that the large $n$ limit of super (or ordinary) Yang-Mills
theory is a weakly coupled theory of neutral
massive particles (``glueballs'') with an effective coupling of order $1/n^2$.
If we represent the particles of the effective theory by fields $B_i$, then
their effective Lagrangian is something like
\eqn\bubuo{L_{eff}=n^2\int d^4x \,\,{\cal L}(B,\nabla B,\nabla^2 B,\dots)}
with corrections of relative order $1/n^2$.  Thus in particular, large $n$ is 
the
regime in which this theory can be treated semiclassically.

If such a theory has several vacua, 
one would at first expect domain walls connecting them to be described as
 soliton-like classical
solutions, in which the fields $B_i$ are independent of the time and two  
spatial 
coordinates, but are non-trivial functions of the third spatial coordinate, 
say 
$x^3$.  The idea would be to find a classical solution in which the $B_i$ 
approach 
one vacuum state for $x ^3\to -\infty$ and another vacuum state for 
$x^3\to +\infty$.

Such a classical solution would have a tension, or energy per unit area,
of order $n^2$.  However, there are cogent reasons to believe that the super
QCD domain wall has a tension that is of order $n$.    
This follows, as we will see,
from the BPS formula for the tension of the domain wall, which
is believed to be BPS-saturated \dvali.

Note that in $N=1$ supersymmetry
in four dimensions, particles cannot be BPS-saturated, as there is no central
charge for particles in the supersymmetry algebra, but either strings
or domain walls can be, since central charges can appear after 
compactification.
Super Yang-Mills theory does not have a central charge for the QCD string (it 
hardly
could as these strings can annihilate in groups of $n$), but it does have
a central charge for domain walls.

In general, in four-dimensional $N=1$ theories, the central charge governing
domain wall tension $T_D$ in a sector with a domain wall interpolating between 
two vacua $a$ and $b$ is the superpotential difference $W(b)-W(a)$.  One way 
to 
prove this is to compactify  the directions transverse to the domain wall
on a very large two-torus of area $A$.  
The theory then reduces to an effective two-dimensional theory, in which
\ref\wittenolive{D. Olive and E. Witten,  ``Supersymmetry Algebras That Include
Central Charges,'' Phys. Lett. {\bf 78B} (1978) 97.} 
the central charge in a sector of states
that interpolate between distinct vacua at spatial infinity is the 
superpotential
difference between the two vacua.  In the two-dimensional theory,
a BPS-saturated domain wall becomes a particle of mass $AT_D$ (or at least
very nearly $AT_D$ in the large $A$ limit).  This particle is BPS-saturated
(or very nearly so for large $A$) so its mass is the absolute value of the
superpotential difference.  Since
the superpotential difference in two dimensions is $A(W(b)-W(a))$ (as the
two-dimensional superpotential is $A$ times the four-dimensional one), the
tension of a BPS-saturated domain wall in four dimensions is $|W(b)-W(a)|$.

If we normalize the super QCD action in the customary way as
\eqn\ilpo{L={1\over 4g^2}\int d^4x 
\Tr\left(F_{IJ}F^{IJ}+\bar\lambda i\Gamma\cdot D
\lambda\right),}
then the large $n$ limit is made by taking $n$ to infinity with
\eqn\komo{\tilde g^2=g^2n}
kept fixed.  It can be shown \ref\ki{K. Intriligator and N. Seiberg,
``Lectures On Supersymmetric Gauge Theories And Electric-Magnetic Duality,''
Nucl. Phys. Proc. Suppl. {\bf 45BC} (1996) 1, hep-th/9509066.}
that the superpotential in any vacuum of this theory is\foot{This
superpotential can be measured as the central charge in a sector with
a domain wall, as discussed above, or by coupling to gravity, whereupon
the superpotential becomes observable.} 
\eqn\jobbo{W=n\langle \Tr \lambda\lambda\rangle.}
On the  other hand, if chiral symmetry breaking is a feature of the leading
large $n$ approximation, then the gluino condensate
$\langle \Tr \lambda\lambda\rangle$, like the expectation value of
any operator that is defined as the trace of a product of elementary fields,
is of order $n$.  
\foot{As M. Shifman has explained, this $n$ dependence and hence the hypothesis
that the gluino condensation is part of the leading large $n$ approximation
can be confirmed using the fact that the gluino condensate of this
theory is actually exactly calculable  \shifman.
Note that if gluino condensation were {\it not} part of the 
leading large $n$ approximation, then $\langle\Tr\lambda\lambda\rangle$
would be smaller than $n$ for large $n$, and the problem raised in the text
would be more severe.} 
So the {\it value} of the superpotential in any given
vacuum is of order $n^2$.  The vacua differ from each other, however,
by a chiral rotation, as a result of which, in the $j^{th}$ vacuum,
one has
\eqn\govo{\langle\Tr\lambda\lambda\rangle_j\sim Cn\exp(2\pi i j/n),}
with $C$ independent of $j$ and of order $\Lambda_{QCD}^3$.   
This   formula, when inserted in \jobbo, shows that while $W$ is of order
$n^2$ in any given vacuum, the {\it differences} between the values of $W$
in neighboring vacua are only of order $n$.  Hence the tension in a 
BPS-saturated
domain wall is only of order $n$.

So the QCD domain wall is not a soliton in the effective large $n$ theory,
an object whose tension would be of order $n^2$.  What is it? 
If we regard  $\lambda= 1/n$ as the string coupling constant of the
QCD string, then we are looking for a nonperturbative object with
a tension of order $1/\lambda$, rather than a conventional soliton
with a tension of order $1/\lambda^2$.  In critical string theory, such objects
have been interpreted as $D$-branes (for a review see \ref\polch{J.
Polchinski, S. Chaudhuri, and C. V. Johnson,
``Notes On $D$-Branes,'' hep-th/9602052, J. Polchinski, ``TASI Lectures
On $D$-Branes,'' hep-th/9611050.}).  
Might the super Yang-Mills domain  wall be a $D$-brane
of the QCD string?  In critical string theory,
a  $D$-brane is an object on which elementary strings
can end.  A $D$-brane in large $n$ QCD would presumably be an object on
which the QCD string can end.

In what follows, we aim to clarify some of these issues.
We will explain how to describe a  domain wall in terms of branes and give
the criterion for such a domain wall to be BPS-saturated.  
Then we will argue  that 
a QCD string really
can end on a super Yang-Mills
 domain wall.  
Finally, we will tie up the loose ends by showing how to calculate the
superpotential of an $M$-theory brane configuration and then
showing that the brane configurations found in section 2 do have a non-zero
superpotential and therefore, via \jobbo, a non-zero gluino condensate.

A puzzle analogous to the one just explained
arises in the supersymmetric ${\bf CP}^n$ model in two dimensions,
where a solitonic domain wall in the large $n$ effective theory would
have a mass of order $n$, but the BPS formula indicates that the mass is
actually of order 1.  In that case, the resolution of the problem
\ref\olderwitten{E. Witten, ``Instantons, The Quark Model, And The
$1/N$ Expansion,''  Nucl.Phys. {\bf B149} (1979) 285.
} is that the domain
wall is not a soliton but is one of the underlying elementary particles.
This is somewhat similar to the fact that in super Yang-Mills theory
the domain wall turns out to be, in a sense, a brane rather than a soliton.
Also, one is reminded of the fact that baryons in the $1/n$ expansion
have masses of order $n$ and appear as solitons in the meson or open string
sector of the large $n$ effective theory \witten.  From a contemporary point
of view,  
one might wonder whether a baryon should be interpreted as
a Dirichlet zerobrane rather than as an open string soliton.  The question,
however, may not be completely well-defined; the example of Type I
fivebranes (which for large scale size are open-string, Yang-Mills solitons
and for small scale size are $D$-branes) shows that solitons of the open-string
sector, unlike closed-string solitons, can be continuously connected to 
$D$-branes.

\subsec{The Domain Wall}

The actual construction of the domain wall is straightforward in concept;
solving the equations is another matter (beyond the scope of the present 
paper).
The domain wall is a physical situation that for $x^3\to -\infty$ looks
like one vacuum of the theory and for $x^3\to +\infty$ looks like another 
vacuum.
Here $x^3$ is one of the three spatial coordinates in $\R^4$.  Meanwhile
the physics should be independent of the time $x^0$ and of the other
two spatial  coordinates $x^1,x^2$.  

Such a physical situation can be described simply as an $M$-theory
fivebrane that interpolates between the fivebranes used to describe the
vacuum states at the two ends.  The vacuum states were described by fivebranes
of the form $\R^4\times \Sigma$, where $\Sigma$ was a Riemann surface embedded
in $\W=\R^5\times \S^1$.  For the domain wall we instead use a fivebrane of the
form $\R^3\times S$, where $\R^3$ is parametrized by $x^0,x^1,x^2$,
and $S$ is a three-surface in the seven manifold $\tilde \W=\R\times \W$
(the copy of $\R$ here being the $x^3$ direction).  Near $x^3=-\infty$,
$S$ should look like $\R\times \Sigma$, where $\Sigma$ is the familiar
Riemann surface defined by $w=\zeta v^{-1}$, $v^n=t$.  Near $x^3=+\infty$,
$S$ should look like $\R\times \Sigma'$, where $\Sigma'$ is the Riemann
surface of an ``adjacent'' vacuum, say defined by $w=\exp(2\pi i/n) \zeta 
v^{-1}$,
$v^n=t$.  These are the only ``ends'' of $S$.

The condition for the domain wall defined by $\R^3\times S$ to be BPS-saturated
is simply that $S$ should be a supersymmetric three-cycle in the sense of 
\refs{
\becker,\morrison}.
In other words, the Riemannian manifold $\tilde \W$, being flat, is a
special case of a manifold of $G_2$ holonomy.  There are unbroken 
supersymmetries
in $M$-theory with an $\R^3\times S$ fivebrane ($S$ being a three-cycle in
$\tilde\W$) 
if and only if $S$ obeys
the conditions of unbroken supersymmetry, that is, the conditions for
a supersymmetric three-cycle.  In view of \refs{\dvali, \vafa}, it seems
highly plausible that such an $S$ exists with the asymptotic behavior that
we have stated.  

Now recall from eqn. \hungo\  that both $\Sigma$ and $\Sigma'$ are invariant 
under a 
$U(1)$ symmetry generated $U$ by $n\partial/\partial t-v\partial/\partial v
+ w\partial/\partial w$.  The question then arises
of whether $S$ is invariant under $U$  or only asymptotically so.  For our
domain wall to be in the universality class of supersymmetric
QCD, $S$ must be invariant
under this symmetry.  The reason is that in supersymmetric QCD, $U$, since it
acts trivially in the field  theory, acts trivially on all degrees of freedom
that propagate on the domain wall.  But if the symmetry is spontaneously
broken on the domain wall, it certainly acts non-trivially in the effective
theory on the domain wall.  Note that as the domain wall is $2+1$-dimensional,
we do not have the option that was available in section 3.2 in the 
superficially
similar case of the QCD string.  The string world-volume is $1+1$-dimensional.
In $1+1$ dimensions, a symmetry that is broken classically will be restored
quantum mechanically at sufficiently big distances, perhaps with generation
of a mass gap.  In $2+1$ dimensions,
this is not necessarily so and will in fact not be so in a region in which
a semiclassical treatment by branes is valid.

\bigskip\noindent{\it $D$-Brane For The QCD String?}

So we have described the domain wall as a brane at least from the point
of view of $M$-theory -- and certainly not as a soliton in an effective 
glueball
theory.  But would a low energy QCD observer think of the domain
wall as a $D$-brane?  Can QCD strings end on the domain wall?
We will now prove that they can.

We recall that the QCD string is represented by a one-brane $C$ in $\W$ that
interpolates, at $t=t_0$ (for some fixed $t_0$) from $v=t_0^{1/n}$  to
$v=t_0^{1/n}e^{2\pi i/n}$.  $C$ does not lie in $\Sigma$, but its endpoints do.
If (keeping the endpoints of $C$ in $\Sigma$ throughout the homotopy) $C$ could
be deformed to a one-brane in $\Sigma$, then $C$ would represent the zero
element of $H_1(\W/\Sigma,\Z)$ and the QCD string would be unstable.

Now we want to analyze the stability of $C$ in the field of a domain wall.
For this purpose, by fixing a large negative value of $x^3$, we regard
$C$ as a one-brane in $\tilde\W=\R\times \W$.  
Thus $C$ is described now by $x^3=-c$ ($c$ a large constant), $t=t_0$, with 
$v$ 
still interpolating from $t_0^{1/n}$ to $t_0^{1/n}e^{2\pi i/n}$.
To show that the QCD string
can  end in the field of the domain wall, it suffices to show that $C$
can be deformed, while keeping its ends on $S$, to a one-brane that lies 
entirely
in $S$.  

For this it suffices to find a one-brane $\eta$ that lies entirely on
$S$, has the same endpoints as $C$, and has the property that $t=t_0b$ where
$b$ is everywhere real and positive.  One can then without changing the
endpoints of $\eta$ make a homotopy from $\eta$ to a one-brane $\eta'$ with
 $b=1$ (that is $t=t_0$).  $\eta'$ and $C$ are two paths in $v-w$ space
with the same endpoints and with $t $ fixed at $t_0$; since $v-w$ space
is topologically trivial (contractible),  $\eta'$ and $C$ are
homotopic by a homotopy that keeps the boundaries fixed.  So $C$ is
equivalent to $\eta$ by such a homotopy, as desired.

In proving existence of $\eta$, we may as well take $t_0$ real and positive
and $\zeta=1$.  This does not affect the topological
question, and 
will slightly simplify the formulas.

We set $\rho=\ln|t|$.  We construct $\eta$ as a path in $S$ with the properties
that (a) $t$ is always real and positive on $\eta$; (b) $\eta$, though not
a closed path on $S$, projects if one forgets all variables except $\rho$
and $x^3$, onto a closed circle at infinity in the $x^3-\rho$
plane.

The projection of $\eta$ to the $x^3-\rho$ plane can be described explicitly as
a closed curve that is built by joining together four pieces.
(1) Start near $x^3=-\infty$, $\rho=\infty$. 
In a vacuum with $vw=1$, vary $\rho$ from
$\infty$ to $-\infty$.  (2) Near $\rho=-\infty$, vary $x^3$ from
$-\infty$ to $+\infty$, going to a vacuum with $vw= e^{2\pi i/n}$. 
(3) Near $x^3=+\infty$, vary $\rho$ from $-\infty$ back to $+\infty$.
(4) Finally, near $\rho=+\infty$, vary $x^3$ from $\infty$ back to $-\infty$,
going back to the starting point.

In step (1), we begin at very large $v$ and with 
$w$ near zero; this corresponds to being very near
$(v,w)=(a,0)$ with very large positive  $a$.
We then, while remaining on $S$,
 interpolate to $\rho=-\infty$, which means small $v$ and large $ w $.
This is done
at  $x^3=-\infty$, so we are on $\Sigma$ with  $v^n=t=e^\rho$, and also
  with  $vw=1$.  After varying $\rho$ in this way, one ends
up very near $(v,w)=(0,a)$ with large positive $a$.  
In step (2), $v$ and $w$ remain fixed.
In step (3), one starts at large $w$, small $v$, and by varying $\rho$
one interpolates to small
$w$, large $v$, on a curve with $v^n=e^\rho=w^{-n}$, 
$vw= e^{2\pi i/n}$.   This brings us to  very near
$(v,w)=(ae^{2\pi i/n},0)$ with
$a$ real and positive.
In step (4), $v$ and $w$ remain fixed again.
Thus, the one-brane $\eta$,  starting very near  $(a,0)$, ends
very near $(ae^{2\pi i/n},0)$, and  hence (for $t_0=a^n$) has the same
endpoints as $C$.  Existence of such an $\eta$, lying entirely on $S$
and with $t$ always real and positive, completes the proof that the
QCD string can end on the domain wall.

By further consideration of the curve $\eta$, it can be proved that if
$S$ is invariant under the $U(1)$ symmetry (as it must be to agree with QCD),
then there is a point in $S$ at which $v=w=0$.  In other words, there is a 
point
in $S$ at which the chiral symmetries are all restored.

\bigskip\noindent{\it Heuristic Interpretation}

S.-J. Rey \rey\
has suggested an intuitive interpretation
\foot{I would like to thank him for giving me permission to summarize the
argument here.} of this result in
terms of 't Hooft's concept of oblique confinement \ref\oblique{G. 't Hooft,
``Topology Of The Gauge Condition And New Confinement Phases In Nonabelian
Gauge Theories,'' Nucl. Phys. {\bf B190} (1981) 455,
``The Topological Mechanism For Permanent Quark Confinement In A Nonabelian
Gauge Theory,'' Phys. Scripta {\bf 25} (1982) 133.}
According to this idea, QCD confinement arises from condensation of
somewhat elusive ``QCD monopoles.''  More generally, there are $n$ possible
confining phases, the condensed object being a dyon, that is a bound state of
a QCD monopole and $k$ quarks, for $0\leq k\leq n-1$.\foot{Notice that the
quarks in question, like the QCD monopoles themselves, 
are somewhat elusive, since the
theory of pure super Yang-Mills theory without chiral multiplets
does not have dynamical quarks as elementary fields.}

If one adiabatically increases the QCD theta angle by $2\pi$, analogy
with the abelian case \ref\thetacharge{E. Witten, ``Dyons Of
Charge $e\theta/2\pi$,'' Phys. Lett. {\bf 86B} (1979) 283.}
suggests that a monopole picks up an electric charge and becomes a dyon,
and more generally that a bound state of a monopole with $k$ quarks is
transformed to a bound state with $k+1$ quarks. Thus the $k^{th}$ vacuum
is adiabatically transformed to the $k+1^{th}$ in increasing $\theta$ by 
$2\pi$.

The pure   super Yang-Mills theory has an anomalous $U(1)_R$ symmetry,
broken by the anomaly down to $\Z_n$.  As a result of this, a shift
in the $\theta$ angle by $\theta\to\theta+\alpha$
can be absorbed in a chiral rotation that acts on the gluino bilinear by
$\Tr\lambda\lambda\to e^{2\pi i\alpha/n}\Tr\lambda\lambda$.
In particular, a $2\pi$ shift in $\theta$ multiplies the gluino
bilinear by $e^{2\pi i/n}$, giving the discrete chiral symmetry
whose spontaneous breaking leads to domain walls.

A domain wall thus separates a state with a given value of $\theta$ from
a state in which $\theta$ is shifted by $2\pi$.
Therefore, according to the notion of oblique confinement,
a domain wall separates a QCD phase in which the condensed
object is a monopole bound to $k$ quarks from a phase in which the
condensed object is a monopole bound to $k+1$ quarks.

Now Rey's intuitive observation is that by combining a composite
of an antidyon on one side of the domain wall (an antimonopole bound to $k$
antiquarks) with a dyon on the other side of the domain wall (a monopole
bound to $k+1$ quarks), one can make a free quark in the domain wall.
Thus the domain wall should support excitations that behave
as though they are in the fundamental representation of $SU(n)$, even though
there are no such fields in the original super Yang-Mills Lagrangian.
But if the domain wall contains quarks, it should be possible for a QCD
string to terminate on a domain wall.  This then is the proposed intuitive
explanation for the ability of a string to end on a domain wall.

\subsec{The Superpotential}

Finally, we will show how to compute directly the superpotential of
an $M$-theory brane configuration.  This will enable us to show from the
brane picture that the super Yang-Mills vacua have a non-zero superpotential,
and hence in particular a gluino condensate.  The method of calculating the
superpotential is also likely to have other applications.

Consider in general $M$-theory compactification on $\R^4\times X \times \R$
 where $X$ is a Calabi-Yau threefold.  We want
to consider a situation in which in spacetime there are fivebranes of the
form $\R^4\times \Sigma$, $\Sigma$ being a two-dimensional real surface in $X$.
The field theory on the brane then is a  
model in $\R^4$ that has global $N=1$ supersymmetry  if $\Sigma$ is
a complex curve in $X$, and no unbroken supersymmetry at all otherwise.
(One gets global rather than local supersymmetry because eleven-dimensional
gravitons and gravitinos are not localized on the brane but propagate in a 
larger
number of non-compact dimensions.
If $X$ has a sufficient amount of non-compactness, one can replace $\R^4\times
X\times \R$ by $\R^4\times X\times \S^1$, and still get a model with
global supersymmetry.) 

For some purposes, it is convenient to think of $\Sigma $ as an abstract 
surface with a map $\Phi:\Sigma\to X$.  From this point of view, one can
introduce local real coordinates $\lambda^a$, $a=1,2$, on $\Sigma$, and local
complex coordinates $\phi^i$, $i=1,\dots,3$ on $X$, and describe $\Phi$
via functions $\phi^i(\lambda^a)$.  These functions are chiral superfields
from the four-dimensional point of view.  In this description, the 
reparametrizations
of the $\lambda^a$ should be regarded as gauge symmetries.

We wish to compute the superpotential $W$ as a function of $\Sigma$.  It should
have the following properties. (1) $W$ should be a holomorphic function of 
$\Sigma$.
(In terms of the last paragraph, this means that $W$ is a holomorphic function
of the chiral superfields $\phi^i(\lambda^a)$ invariant under 
reparametrizations of
$\Sigma$.)
  (2) $W$ should have a critical point
precisely when $\Sigma$ (or  $\Phi(\Sigma)$) is a holomorphic
curve in $X$, this being the condition for unbroken supersymmetry.
Note that as we are doing global supersymmetry, $W$ need be defined only
up to an overall additive constant.

It is not difficult to describe the functional that obeys these 
conditions.\foot{
Closely related issues were discussed by S. Donaldson in a lecture at
the Newton Institute at Cambridge University, November, 1996.}
Let $\Omega$ be the holomorphic three-form of $X$.  We assume first that 
$\Sigma$
is compact, and  (unrealistically for the sake of most applications) that the
homology class of $\Sigma$ is zero so that $\Sigma$ is the boundary of a
three-manifold $B$.  The superpotential is then
\eqn\supis{W(\Sigma)=\int_B \Omega.}
In the description of $\Sigma$ by chiral superfields $\phi^i(\lambda^a)$, this
means that the variation of $W$ in a change in $\Sigma$ is
\eqn\upis{\delta W=\int_\Sigma\Omega_{ijk}\delta \phi^i d\phi^j\wedge d\phi^k,}
where of course $d=\sum_{a=1,2}d\lambda^a \,\partial/\partial\lambda^a$.
From this formula, it is straightforward to show the two desired properties of
$W$. The fact that \upis\ is proportional to $\delta \phi^i$ with no
$\delta\overline\phi^i$ means that $W$ is holomorphic.  Moreover, $W$  
is stationary for and only for $\Sigma$ a holomorphic curve, because
this condition is equivalent to $d\phi^j\wedge d\phi^k=0$.

This generalizes straightforwardly
 to the more typical case that the homology class of $\Sigma$
is non-zero.  One picks a fixed $\Sigma_0$ in the homology class of $\Sigma$, 
and picks a three-manifold (or more generally a three-cycle)
$B$ with boundary $\Sigma-\Sigma_0$ (that is,
the boundary of $B$ consists of the union of $\Sigma$ and $\Sigma_0$, which
appear with opposite orientation), and one defines
\eqn\pupil{W(\Sigma)-W(\Sigma_0) =\int_B\Omega.}
This defines $W(\Sigma)$ up to an additive constant $W(\Sigma_0)$.

There are actually
two reasons for the indeterminacy of this additive constant.
One is the arbitrary choice of $\Sigma_0$.  
In addition, if $H_3(X,\R)$ is non-zero, then there are different possible 
choices for the homology class of $B$; changing this class will shift 
$W(\Sigma)$
by a constant.  Note finally that if $H_3(X,\Z)$ is non-zero
and the space of possible $\Sigma$'s is not
simply-connected, then $W(\Sigma)$ may change by an additive constant in
going around a loop in the space of $\Sigma$'s.  Thus, $W$ is really only
defined on the universal cover of the space of $\Sigma$'s.  In our actual
application, $H_3(X,\Z)=0$ and the subtleties mentioned in the present 
paragraph
do not arise.

In the above, we have assumed that $\Sigma$ is compact.  If not, one should
require that $\Sigma$ and $\Sigma_0$ have the same asymptotic behavior at
infinity, and that $B$ is ``constant'' at infinity.  A slight generalization
of the above is that $\Sigma_0$ (and for that matter $\Sigma$) need not
actually be submanifolds of space-time.  One can use ``cycles,'' which for
our purposes we can take to mean that one can regard $\Sigma_0$ as an abstract
real oriented  surface with a differentiable
map $\Phi_0:\Sigma_0\to X$.  The map need {\it not}
be an embedding.  Likewise, we can consider $B$ to be an abstract oriented
three-manifold (of boundary $\Sigma-\Sigma_0$) with a map $\Phi_B:B\to X$
(again, not necessarily an embedding).  The restriction of $\Phi_B$ to $\Sigma$
should give the surface in $X$ for which we wish to evaluate the 
superpotential,
and the restriction of $\Phi_B$ to $\Sigma_0$ should coincide with $\Phi_0$.
In this context the definition of the superpotential is
\eqn\urry{W(\Sigma)-W(\Sigma_0)=\int_B\Phi_B^*(\Omega).}

\bigskip\noindent{\it Computation Of $W$}

Now we want to actually compute $W$ for the specific brane $\Sigma$
that represents the super Yang-Mills theory vacuum.  We recall that $\Sigma$
is defined by the equations $v^n=t$, $w=\zeta v^{-1}$.  

By computing $W$, we really mean  determining its $\zeta$-dependence and
in particular (since the different vacua are permuted by $\zeta\to\zeta 
e^{2\pi i/n}$) comparing the values of $W$ for the different vacua.
However, according to \pupil\ or \urry, 
to define $W$ we must
 pick a base-point $\Sigma_0$ in the space of possible $\Sigma$'s.
Ideally, we would like to pick a chirally-symmetric $\Sigma_0$ and
define $W(\Sigma_0)=0$, so as to get a chiral-invariant definition of $W$.
Since it is at best awkward to find a chiral-invariant $\Sigma_0$, we will
proceed by a slight variant of this. (Note
that $H_3(\W,\Z)=0$, so there is no indeterminacy in $W$ coming from periods
of $\Omega$.)

In doing the computation, it will be useful to use the formulation of
equation \urry\ in which $\Sigma_0$ and $B$ are not 
embedded in $\W=\R^5\times \S^1$,
but are abstract manifolds with maps to $\W$. 
  We introduce a complex variable
$r$ and take $\Sigma_0$ to be the complex $r$-plane with the origin deleted.
We write $r=\exp(\rho+i\theta)$, with $\rho$ and $\theta$ real, and
pick an arbitrary smooth function $f$ of a real variable such that
$f(\rho)=1$ for $\rho>2$ and $f(\rho)=0$ for $\rho<1$.  Then we define
the map $\Phi_0:\Sigma_0\to \W$ by\foot{This map fails to be an embedding
in the region $-1\leq \rho \leq 1$ where $f(\rho)=f(-\rho)=0$.  It is hard
to avoid this for a chiral-invariant $\Sigma_0$.  That is why we have 
introduced
the slightly more abstract formalism with cycles rather than submanifolds.}
\eqn\iccob{\eqalign{t & = r^n \cr
                    v & = f(\rho) r\cr
                    w & = \zeta f(-\rho) r^{-1}.\cr}}
Note that $\Sigma_0$ is asymptotic at infinity to $\Sigma$.
$\Sigma_0$ is close enough to being invariant under the $\Z_n$ chiral
symmetry to that a chirally-invariant definition of $W$ will be obtained
by setting $W(\Sigma_0)=0$.
This will be shown by a slightly technical argument which will occupy
the rest of this paragraph.\foot{An error
in an earlier version of this argument
was pointed out by M. Schmaltz.}
Let $\Sigma'$ be the chirally rotated version of $\Sigma_0$ with
$\zeta$ replaced by $\zeta e^{2\pi i/n}$. 
I will show that $W(\Sigma_0)=W(\Sigma')$, which shows that to achieve 
a chirally-symmetric definition of $W$, one must fix the additive constant
in the definition of $W$  so that $W(\Sigma_0)=0$.  
To show that $W(\Sigma_0)=W(\Sigma')$, note that 
$\Sigma_0$ and $\Sigma'$ are
described by two different but related maps of the $  r$-plane to $\W$.
Consider a (non-holomorphic) reparametrization of the $r$-plane
defined by $\rho\to \rho$, $\theta\to\theta+b(\rho)$, 
where $b$ is any continuous function with $b(\rho)=0$ for $\rho\geq 1$, and
$b(\rho)=-2\pi/n$ for $\rho\leq -1$.   This reparametrization maps
$\Sigma_0$ into $\Sigma'$ except in the region  $-1\leq\rho\leq 1$,
and in that region $v$ and $w$ are both mapped to zero.  The region
$-1\leq\rho\leq 1$  of the $r$-plane is an annulus; by gluing together
two copies of this annulus along the boundary circles at $\rho=\pm 1$ 
we make a two-torus $T$. 
The difference $\Sigma_0-\Sigma'$ can be represented by  a two-cycle
consisting of a map $\phi$
of $T$ to $\W$ whose image is entirely at $v=w=0$.
Thus, $\phi$  
is really a map  of $T$ to the punctured $t$-plane (with $t=0$ omitted);
we call the punctured $t$-plane $V$.  Since $H_2(V,\Z)=0$, we can  find
a three-manifold $B$, of boundary $T$, and a map $\Phi:B\to V$ whose
restriction to $T$ coincides with $\phi$.  As $\Phi(B)$ is at $v=w=0$,
we have $\int_B\Phi^*\Omega=0$, so $0=W(T)=W(\Sigma_0)-W(\Sigma')$, as was to 
be proved.

We take $\Sigma$ to be again the $r$-plane, mapped to $\W$ by similar
formulas, but with $f$ replaced by 1:
\eqn\niccohb{\eqalign{ t & = r^n \cr
                       v & = r   \cr
                       w & = \zeta r^{-1}. \cr}}

To compute $W(\Sigma)$, we need to pick a three-manifold with boundary
$\Sigma-\Sigma_0$, and an appropriate map $\Phi_B:B\to  \W$.
We take $B$ to be the product of the punctured $r$-plane (with the origin
deleted) and the unit interval $I$, parametrized by a real variable $\sigma$,
$0\leq\sigma\leq 1$. We identify $\Sigma$ with the boundary at $\sigma=1$
and $\Sigma_0$ with the boundary at $\sigma=0$.  We introduce a smooth bounded
function
$g(\rho,\sigma)$ with $g(\rho,1)=1$, $g(\rho,0)=f(\rho)$, and 
$g(\rho,\sigma)=1$
for $\rho>2$.  The map $\Phi_B:B\to \W$ can then be defined by
\eqn\milcob{\eqalign{ t & = r^n \cr
                      v & =g(\rho,\sigma)r \cr
                      w & = \zeta g(-\rho,\sigma)r^{-1}.\cr}}
The definition
\eqn\pilcob{W(\Sigma)=\int_B\Phi_B^*(\Omega)}
of the superpotential now becomes
(with $\Omega = R \,dv \wedge dw\wedge dt/t$ and $dt/t=n\,dr/r$)
\eqn\rufo{W(\Sigma)=Rn\int_B {dr\over r}\wedge dv\wedge dw.}
This is more explicitly
\eqn\ruffo{W(\Sigma)=iRn\zeta
\int_0^1d\sigma\int_0^{2\pi} d\theta\int_{-\infty}^\infty
d\rho  \left({\partial g_+\over \partial\sigma}{\partial g_-\over\partial\rho}
-{\partial g_+\over\partial \rho}{\partial g_-\over\partial\sigma}\right),}
with $g_\pm(\rho,\sigma)=g(\pm \rho,\sigma)$.  
The integral over $\theta$ gives a factor of $2\pi$.  The  integral
over the remaining variables
is the integral over $\R\times I$ (the two factors being parametrized
by $\sigma$ and
$\rho$) of the exact two-form $d(g_+dg_--g_-dg_+)$.  Integrating by parts
and picking up surface terms at $\rho=\pm \infty$, this integral equals 2.
So finally\foot{As we have not been
careful about an overall multiplicative constant in the definition of
$\Omega$, the following formula (and others deduced from it later)
are uncertain by such a universal constant.} 
\eqn\ubbub{W(\Sigma)=4\pi i Rn\zeta.}

In particular, the theory has a gluino condensate, 
$\langle\Tr\lambda\lambda\rangle
= 4\pi iR\zeta$.  We also can now get some insight about the large $n$ limit.
Since the QCD string had a tension of order $|\zeta|^{1/2}/n$, a smooth
large $n$ limit of the QCD string tension required $\zeta\sim n^2$.
To make $W$ of order $n^2$, we hence need $R\sim 1/n$.

Ordinarily, Kaluza-Klein excitations carrying momentum in the $x^{10}$ 
direction
have energies of order $2\pi/R$, so the result $R\sim 1/n$ would appear at 
first
sight to show that these modes decouple in the large $n$ limit.  We must,
however, be careful in drawing such a conclusion for modes that are localized
on the fivebrane $\R^4\times \Sigma$.  Because $\Sigma$ wraps $n$ times
around the $x^{10}$ circle, if the brane can be treated semiclassically one
would expect such modes to 
see an effective radius of order $nR$,
that is, of order 1.  Thus, it appears that this sort of Kaluza-Klein 
excitation
on the fivebrane survives for $n\to \infty$.   Of course, we must decouple
such modes (along with others)
if we are to get super Yang-Mills quantitatively, and not just
a theory that shares many qualitative properties with super Yang-Mills theory.
What is involved in decoupling such modes will be the subject of the next 
section.

\nref\susskind{
J. M. Maldacena and L. Susskind, ``$D$-Branes And Fat Black Holes,''
Nucl. Phys. {\bf B475} (1996) 679, hep-th/9604042.}
\nref\motl{L. Motl, ``Proposals On Non-Perturbative Superstring Interactions,''
hep-th/9701025.}
\nref\bankseiberg{T. Banks and N. Seiberg, ``Strings From Matrices,''
hep-th/9702187.}
\nref\dvv{R. Dijkgraaf, E. Verlinde, and H. Verlinde,
``Matrix String Theory,''
hep-th/9703030.}
The mechanism whereby the Kaluza-Klein modes on the fivebrane survive for
$n\to \infty$ -- a brane wrapped $n$ times around a circle with radius of order
$1/n$ sees an effective radius of order $1$ -- is highly reminiscent of
recent results in black hole theory \susskind\
and matrix string theory \refs{\motl - \dvv}.

Since the domain wall tension is the magnitude of the
difference between the values of $W$ in
different vacua, \ubbub\ leads to a precise formula for the domain wall tension
$T_D$:
\eqn\immo{T_D=4\pi nR\left|\zeta(1-e^{2\pi i/n})\right|.}

\newsec{Interpretation}

It is surprising to be able to understand semiclassically properties
like those we have explored in this paper -- confinement, chiral symmetry
breaking, and the fact that the domain wall behaves as a $D$-brane.
In this section, we will demystify this success a  bit, by showing
that it depends on the fact that the model we have analyzed,
while evidently in the same universality class as the super Yang-Mills 
four-dimensional field theory, is actually a different theory.

Two of our main results are the estimates for the tension $T$ of the QCD string
and the tension $T_D$ of the domain wall.  In order of magnitude, in
units in which the eleven-dimensional Planck scale is one, one has
\eqn\hobo{ T\sim {|\zeta|^{1/2}\over n}.}
One also has a precise BPS formula \immo\ for the domain wall tension, which
in order of magnitude is
\eqn\nobo{T_D=  R|\zeta|.}

Since \nobo\ is a BPS formula, it holds for all values of the parameters
$R$, $\zeta$.  The same cannot be said for \hobo, which is valid only to
the extent that the QCD string can be treated semiclassically.  We recall
that \hobo\ has a simple origin.  Our fivebrane wrapped $n$ times around
the $\S^1$ factor in $\W=\R^5\times \S^1$; the distance between neighboring
branches was of order $|\zeta|^{1/2}/n$, and to the extent that fluctuations
in the string position are not important, this is the string tension.
Such fluctuations will be suppressed if $|\zeta|^{1/2}/n$ and $R$ are  large
(in eleven-dimensional Planck units).  But in that case the string tension
is much greater than the eleven-dimensional gravitational scale, and one
may wonder whether the complexities of $M$-theory are really well-separated
from the four-dimensional super Yang-Mills physics.  After all, in the 
Yang-Mills field theory, quantum fluctuations in the string are believed
to be important -- so important that it has been hard to see the existence
of the string at all, in the continuum quantum Yang-Mills theory.

To address this question further, let us ask whether it is possible
for the brane theory to agree with super Yang-Mills theory for values of the
parameters for which \hobo\ holds.  In super Yang-Mills theory, there is
a mass scale $\Lambda$, and one has $T\sim \Lambda^2$, $T_D\sim n\Lambda^3$.
So \hobo\ and \nobo\ enable us to evaluate the $M$-theory parameters in
terms of $\Lambda$. We get
\eqn\kobo{\eqalign{\zeta& \sim n^2\Lambda^4\cr
                     R & \sim {1\over n\Lambda}.\cr}}

Can this theory be super Yang-Mills theory 
in any region in which \kobo\ holds?  
Kaluza-Klein excitations, with momentum around the $\S^1$ factor in $\W$, 
must decouple
in any limit in which this happens.  It was argued at the end of the last 
section
that such modes have masses of order $1/Rn$, that is of order $\Lambda$.
Thus they do not decouple; their masses are comparable to the QCD scale. 

Thus, our theory is not super Yang-Mills theory, though it is evidently
a close cousin of it.  It is not super Yang-Mills theory because it depends
on two parameters, $\zeta $ and $R$, while super Yang-Mills theory has
only one parameter, $\Lambda$.  The formulas $T\sim |\zeta|^{1/2}/n$,
$T_D\sim R|\zeta|$ make it clear that the theory really depends
independently on $R$ and $\zeta$.  What has apparently happened is that
the brane picture has given us
a surprising generalization of the conventional super Yang-Mills theory,
in which one extra parameter is introduced in a way that preserves 
gauge-invariance and relativity and simplifies the dynamics.  

An analogous statement applies to many recent papers in which brane 
configurations
are used to study various four-dimensional (or three-dimensional)
field theories.  Branes achieve
their simplification by giving us not the familiar four-dimensional field
theories but simpler and for some purposes equivalent cousins.

\bigskip\noindent{\it Compactification From Six Dimensions}

If the theory that we have been studying in this paper is not really
super Yang-Mills theory, then what is it?  To answer this question, it
is helpful to note that intuitively, if we are to try to get super Yang-Mills
theory from the brane configuration, $\Lambda$ should be extremely small
compared to the eleven-dimensional Planck scale.

Thus, in \kobo, $R$ must be very large, and $\zeta$ very small.
The fact that $R$ is large means that the fivebrane is  large in all six
world-volume dimensions.  The fact that $\zeta$ is small means that
the different branches of the fivebrane are very close together.
The fivebrane, as it wraps $n$ around the $\S^1$ factor in $\W$, thus
looks locally like a system of $n$ almost coincident parallel fivebranes.

This strongly suggests that the system should be understood as an unusual
compactification of the six-dimensional $(0,2)$ superconformal field theory 
\uwitten, which can be described in terms of parallel fivebranes \strom.
In this field theory, the separation $\epsilon$ between neighboring
fivebranes is interpreted as a scalar field of dimension two.  There is
therefore a limit in which $\epsilon$ goes to zero, with all six-dimensional
lengths being scaled as $1/\epsilon^{1/2}$.

This is precisely the scaling that we see in \kobo.
In other words, according to \kobo, the separation $\epsilon=|\zeta|^{1/2}/n$
between branes is of order $\Lambda^2$.  Since $\Lambda$ is the QCD mass scale,
this is the expected statement that $\epsilon$ behaves as a field of
 dimension two.  Meanwhile the formula $R\sim 1/n\Lambda$ shows that $R$
has dimension $-1$, as expected in conformal field theory.

The whole setup can now be described more or less precisely as a sort
of exotic compactification of the six-dimensional
$(0,2)$ superconformal field theory.
We start with the six-manifold $V=\R^4\times \R\times \S^1$, where $\R^4$
is to be interpreted as four-dimensional Minkowski space, and $\R$ and $\S^1$
are parametrized by $x^6$ and $x^{10}$.  As usual, we take $R$ to be the
radius of the $\S^1$, and introduce $s=R^{-1}x^6+i
x^{10}$ and $t=e^{-s}$.

  We want to consider the six-dimensional $A_{n-1}$ superconformal
field theory (in other words, the theory that can be realized with
$n$ parallel fivebranes) on the six-manifold $V$.  In this theory,
$v=x^3+ix^4$ and $w=x^7+ix^8$ can be interpreted (along with $x^9$, which
we set to zero asymptotically) as order parameters or scalar fields of
dimension two.   
To be more precise, as there are $n$ fivebranes, these order parameters
are $n$-valued.  Moreover, as the $(0,2)$ superconformal field theory is
not well understood, I only claim that these $n$-valued order parameters
are well-defined when they are large, in units set by some other length
scale (such as $R$) characteristic of a given physical problem.

I claim that the problem studied in the present paper is
equivalent to studying  the $A_{n-1}$ superconformal theory
on $V$ in a situation in which the asymptotic behavior of $v$ for
$x^6\to\infty$ is 
\eqn\ikko{v^n=t, }
with $w$ small,
and the asymptotic behavior for
$x^6\to -\infty$ is
\eqn\hikko{w^n=\zeta t^{-1},}
with $v$ small.  To the extent that the $A_{n-1}$ superconformal theory
can be understood semiclassically via branes, these boundary conditions
lead back immediately to the configuration studied in the present paper.

Conformal invariance of the (0,2) field theory
means that the physics with this sort of compactification is 
invariant under rescaling of $\zeta$ or equivalently of 
$\epsilon=|\zeta|^{1/2}/n$
provided that one keeps
\eqn\polp{y=\epsilon R^2}
fixed.  $y$ is the extra dimensionless parameter not present in ordinary
super Yang-Mills theory.  If $y$ is too large, the candidate QCD string
does not have the correct universality class, as we saw in section 3.2.

In what limit might one recover the conventional four-dimensional
super Yang-Mills theory?  This apparently  must be the limit of $y\to 0$,
which if we go back to the full $M$-theory (rather than the $(0,2)$ 
superconformal
field theory which is a limit of $M$-theory) can be interpreted to mean
that $R$ is small and $M$-theory degenerates to the weakly coupled
Type IIA superstring.  That, after all, is where our discussion started.
Whatever else may happen,  conventional Yang-Mills theory can arise
as a limit of weakly coupled Type IIA superstring theory.

\bigskip\noindent{\it Another Parameter}

Another parameter could be introduced by replacing the $M$-theory
manifold $\R^4\times \W\times \R$, in which we have worked in this paper,
by $\R^4\times \W\times {\bf S}^1$, and judiciously varying the radius of the
new $\S^1$.  In this fashion, one gets a deformation of super Yang-Mills
theory depending on two extra parameters (essentially the radius of the circle
factor in $\W$ and of the new circle) not seen in conventional
four-dimensional super Yang-Mills theory.  By analogy with what has just been 
said,
this theory can be understood in terms of compactification of the 
six-dimensional
non-critical string theory \ref\seibnew{N. Seiberg, 
``New Theories In Six Dimensions And Matrix Description Of $M$ Theory On
${\bf T}^5$ and ${\bf T}^5/{\bf Z}_2$,'' hep-th/9705221.} which generalizes
the six-dimensional $(0,2)$ field theory.

\newsec{The Non-Supersymmetric Case}

To conclude, we would like to argue that the brane method may be relevant
to theories that are not supersymmetric, for instance to ordinary 
four-dimensional
Yang-Mills theory without supersymmetry.  The problem, as we will see,
is not to find the brane configurations but to determine  what aspects of
the Yang-Mills physics can be understood from them.  Also, in the 
nonsupersymmetric case there is an extra potential pitfall.  In going from
the field theory to the brane system, there may in the absence of
supersymmetry be a first order phase
transition, which might make it impossible in principle to learn anything
about Yang-Mills theory by studying the branes.  We must hope that there is
no such first-order transition.

We start in weakly coupled Type IIA superstring theory with $n$ fourbranes
suspended between two parallel fivebranes, a system of $N=2$ supersymmetry.
Just as one can ``rotate'' the fivebranes to
break $N=2$ supersymmetry down to $N=1$ \rotat,
a further generic ``rotation'' breaks
the supersymmetry completely.  This gives the gluinos a bare
mass, but one still has the $SU(n)$ gauge symmetry of $n$ parallel fourbranes.
So in a certain limit, one obtains the conventional, bosonic $SU(n)$ Yang-Mills
theory.

To study this theory via branes, we go to a strongly-coupled limit,
which is $M$-theory on $\R^{10}\times \S^1$, with a fivebrane that
-- as in the $N=1$ case -- wraps $n$ times around the $\S^1$, 
but has a supersymmetry-breaking asymptotic behavior at infinity.  
As before, we write $\R^{10}\times \S^1=\R^4\times \W\times \R$, where $\R^4$ 
is the effective Minkowski space and $\W=\R^5\times \S^1$.
The desired fivebrane will be of the form $\R^4\times \Sigma$, where $\Sigma$
is a two-dimensional surface in $W$.  $\Sigma$ will go to infinity in
certain directions, corresponding to the fact that in the Type IIA description,
there are two non-compact fivebranes.  $\Sigma$ cannot be found by
requiring holomorphy, since (supersymmetry being completely broken) 
$\Sigma$ is not a holomorphic curve in any complex structure on $W$.  $\Sigma$
must be found by minimizing its area subject to a given asymptotic
behavior at infinity.  

Luckily, the techniques for doing so are familiar from the theory of the
classical bosonic string.  $\W$ has a flat Riemannian metric
\eqn\yeqn{ds^2=\sum_{i,j=4,\dots,8,10}\eta_{ij}dx^idx^j=\sum_{i=4,\dots 
,8}(dx^i)^2
+R^2(dx^{10})^2.}
The embedding of $\Sigma$ in $\W$ induces
a metric and therefore a complex structure on $\Sigma$.  We can assume that
$\Sigma$ has genus zero, since this is the case even before the rotation
from $N=1$ to $N=0$, and the genus of a smooth surface does not change under
a generic 
rotation.  Hence, $\Sigma$ as a complex manifold is isomorphic to ${\bf P}^1$,
perhaps with some points deleted.  The number of deleted points is precisely
two, as we want to study a situation with precisely two infinite fivebranes.
Hence we can identify $\Sigma $ with the complex $\lambda$-plane, with the
origin deleted.

The embedding of the $\lambda$-plane in $\W$ is not holomorphic in any
complex structure on $\W$.  However, it is a minimal area embedding.
As in the classical theory of the bosonic string, this means that
the coordinates $x^4,\dots,x^8$ and $x^{10}$ are harmonic functions
on the $\lambda$-plane which  obey the Virasoro constraints
\eqn\murmo{\eta_{ij}{dx^i\over d\lambda}{dx^j\over d\lambda}=0.}
The Virasoro constraints are automatic if the embedding of $\Sigma$ in $\W$
is holomorphic, but as it is we will have to check the Virasoro constraints
by hand.

\def\RL{{\rm Re}\,}
\def\IM{{\rm Im}\,}
In the $N=1$ case, the embedding of $\Sigma$ in $\W$ is $v=\lambda$, 
$w=\zeta\lambda^{-1}$, $t=\lambda^n$.  With $v=x^4+ix^5$, $w=x^7+ix^8$,
$t=\exp(-(R^{-1}x^6+ix^{10}))$, these equations can be written
\eqn\turmo{\eqalign{ x^4&    =\RL \lambda \cr
                     x^5 &   =\IM \lambda\cr 
                     x^7 &   =\RL(\zeta\lambda^{-1})\cr
                     x^8 &   =\IM(\zeta\lambda^{-1}) \cr
                     x^6 &   = -(Rn)\,\, \RL\ln\lambda \cr
                     x^{10}& = -n\,\,\IM\ln\lambda.\cr}}
To rotate this to a nonsupersymmetric configuration that still obeys
the Virasoro constraints, it suffices to do the following.  Combine
$x^4,x^5,x^7,$ and $x^8$ into a real four-vector $\vec A$. 
Generalize the first four equations in \turmo\ to
\eqn\purmo{\vec A=\RL\left(\vec p \lambda +\vec q\lambda^{-1}\right),}
with complex four-vectors $\vec p, \vec q$.  Meanwhile, generalize
the last two equations in \turmo\ to
\eqn\xurmo{\eqalign{x^6 & = -(Rnc)\,\, \RL\ln\lambda \cr
                     x^{10}& = -n\IM\ln\lambda\cr}}
with a real constant $c$.
The Virasoro constraints turn out to be
\eqn\hurmo{\vec p\,^2=\vec q\,^2=0,
\,\,\,~~-\vec p\cdot \vec q +{R^2n^2\over 2}(1-c^2)
=0.}
The condition for unbroken supersymmetry is that in addition $\vec p\cdot \vec 
q=0$,
$c=\pm 1$.  
Simply by picking complex null vectors $\vec p $, $\vec q$ with $\vec p\cdot 
\vec q\not= 0$, and adjusting $c$ appropriately, we get nonsupersymmetric
solutions of \hurmo.  

Clearly, a number of parameters appear in this general description.
An amusing special case is that in which $\vec q$ is the complex conjugate
of $\vec p$ and $c=0$.  In this 
case, $\Sigma$ is embedded as a minimal area surface in a space
of only {\it three} real dimensions (namely the $x^{10}$ direction
and the directions spanned by the real and imaginary parts of $\vec p$).

Rotating the brane and deforming away from $c=1$ does not change the topology
of the situation, so the QCD string still exists in this nonsupersymmetric
theory.  Obviously, our considerations about chiral symmetry breaking
and domain walls do not carry over in this situation.

\bigskip

I would like to thank S.-J. Rey, N. Seiberg,  M. Schmaltz,
M. Shifman, and C. Vafa for helpful 
communications and discussions.
\listrefs
\end